\documentclass[10pt]{article}
\usepackage{lscape}
\usepackage{amsfonts}
\usepackage{amsmath}
\usepackage{amsthm}
\usepackage{amssymb}
\usepackage{graphicx}
\usepackage{textcomp}

\def\be{\begin{eqnarray}}
\def\ee{\end{eqnarray}}
\def\nn{\nonumber}

\def\Tr{{\rm Tr}\,}

\def\l[{\phantom.[}

\def\Dq{\Delta}

\hoffset= - 1.2in         

\begin{document}

\title{\vspace{.1cm}{\Large {\bf  HOMFLY polynomials in representation $[3,1]$
for 3-strand braids}\vspace{.2cm}}
\author{
{\bf A.Mironov$^{a,b,c,d}$}\footnote{mironov@lpi.ru; mironov@itep.ru},
\ {\bf A.Morozov$^{b,c,d}$}\thanks{morozov@itep.ru},
\ {\bf An.Morozov$^{c,d,e}$}\footnote{andrey.morozov@itep.ru},
\ \ and
 \ {\bf A.Sleptsov$^{b,c,d,e}$}\thanks{sleptsov@itep.ru}}
\date{ }
}

\maketitle

\vspace{-5cm}

\begin{center}
\hfill FIAN/TD-08/16\\
\hfill IITP/TH-05/16\\
\hfill ITEP/TH-04/16
\end{center}

\vspace{3.3cm}

\begin{center}
$^a$ {\small {\it Lebedev Physics Institute, Moscow 119991, Russia}}\\
$^b$ {\small {\it ITEP, Moscow 117218, Russia}}\\
$^c$ {\small {\it Institute for Information Transmission Problems, Moscow 127994, Russia}}\\
$^d$ {\small {\it National Research Nuclear University MEPhI, Moscow 115409, Russia }}\\
$^e$ {\small {\it Laboratory of Quantum Topology, Chelyabinsk State University, Chelyabinsk 454001, Russia }}

\end{center}

\vspace{.5cm}

\begin{abstract}
This paper is
a new step in the project of
systematic description of colored knot polynomials started in \cite{MMfing}.
In this paper, we managed to explicitly find
the {\it inclusive} Racah matrix, i.e.
the whole set of mixing matrices in channels $R^{\otimes 3}\longrightarrow Q$
with all possible $Q$, for $R=[3,1]$.
The calculation is made possible by the use of a newly-developed efficient highest-weight method,
still it remains tedious.
The result allows one to evaluate and investigate $[3,1]$-colored polynomials for
arbitrary 3-strand knots, and this confirms many previous conjectures
on various factorizations, universality, and differential expansions.
We consider in some detail the next-to-twist-knots   three-strand family $(n,-1\,|\,1,-1)$
and deduce its colored HOMFLY. Also confirmed and clarified is the eigenvalue hypothesis for the Racah matrices,
which promises
to provide
a shortcut to generic formulas for arbitrary representations.

\end{abstract}

\vspace{.5cm}

\section{Introduction}

Colored knot/link polynomials \cite{knotpols,Con}
are characteristics of knots and links, which factorize into products
for composite knots.
In this sense, the knot polynomials for the {\it prime} knots play the role of {\it prime} numbers,
only in the world of knots, and the most challenging is the question if they
are indeed {\it prime} or there is some additional "more elementary" structure,
which allows one to reconstruct them from some simpler constituents.
Another source of interest to the colored knot/link polynomials is that they are the Wilson loop averages
in Chern-Simons theory \cite{CS}, the simplest of all Yang-Mills theories,
and they can be exactly calculated {\it non-perturbatively},
because they are {\it polynomials} in the variable $q=\exp(g^{-2})$.
This provides a very non-trivial check for our emerging understanding of non-perturbative
methods in quantum field and string theory.
Both these reasons make evaluation of colored knot polynomials extremely important problem,
since it is the first step towards understanding their properties.
Until recently, it was an unachievable task, but
development of theoretical methods in \cite{RT}-\cite{Garou},
combined with the current computer power makes it nearly realistic.

This explains our reasons to make a try, and it is already partly successful.
At the previous stages,
the calculus for totally symmetric and antisymmetric representations was successfully developed
\cite{sympols},
then extended in \cite{Ano21,GJ,twist,nrv,MMMRS,MMMS21} to representation $[2,1]$, and, in the present paper,
we report a new achievement: results for 3-strand polynomials in representation $[3,1]$.
The next big challenge is the first two-hook representation $[4,2]$, this requires a
new serious theoretical advance, but now it seems within reach.
The powerful method needed for that purpose will be described in detail in a separate publication,
here we just mention it in general description of our approach.

\bigskip

It involves the following steps:

\bigskip

{\bf 1.} Define the representation $R$ by a highest weight in the space $[1]^{\otimes R}$.

\bigskip

{\bf 2.} Find the highest weight of representation $Y\in R\otimes R$.

\bigskip

{\bf 3.} Distinguish between $Y$ belonging to the symmetric and antisymmetric squares of $R$.
If a given representation $Y$ appears in these both, we treat $Y_+$ and $Y_-$ as
different representations: they can have there own multiplicities, but they are
never summed.
The reason for this separation of $Y_\pm$ is that the knot polynomials depend on the
eigenvalues of quantum ${\cal R}$-matrices: these are different (in sign) for
$Y_+$ and $Y_-$.

\bigskip

{\bf 4.} Find highest weights of representations $Q \in Y\otimes R$ and $Q \in R\otimes Y$.
These weights $h_Q^l(Y)$ and $h_Q^r(Y)$ (superscripts stand for "left" and "right")
are related by the Racah (mixing) matrix ${\cal U}_Q$:
\be
h_Q^r(Y) = \sum_{Y'\in R\otimes R} {\cal U}_Q(Y,Y')\, h_{Q}^l(Y')
\ee
In the case of non-trivial multiplicities, ${\cal U}$ is defined modulo rotations in the
multiplicity spaces of $Q$, $Y$ and $Y'$, but these rotations leave ${\cal R}$-matrix intact,
and therefore do not affect knot polynomials.
These rotations, however, can be essential for the eigenvalue hypothesis, expressing
the entries of ${\cal U}$ through those of ${\cal R}$, then this freedom should be
somehow fixed.

\bigskip

{\bf 5.} Evaluate the reduced (normalized to unknot) knot/link polynomial for a knot represented by a closure of the 3-strand braid
$B^{(m_1,n_1|m_2,n_2|\ldots)}$  by \cite{MMMkn12}
\be
H_R^{(m_1,n_1|m_2,n_2|\ldots)} = \sum_{Q\in R^{\otimes 3}}
\ \frac{D_Q}{D_R}\cdot
\Tr_Q  \Big\{ {\cal R}_Q^{m_1}{\cal U}_Q {\cal R}_Q^{n_1}{\cal U}^\dagger_Q
{\cal R}_Q^{m_2}{\cal U}_Q {\cal R}_Q^{n_2}{\cal U}^\dagger_Q \ldots\Big\}
\label{3strafla}
\ee
In the following picture $m_1=0,n_1= -2,m_2=2,n_2=-1,m_3=3$:

\bigskip

\unitlength 0.8mm 
\linethickness{1pt}
\ifx\plotpoint\undefined\newsavebox{\plotpoint}\fi 
\begin{picture}(145.5,53)(-30,0)
\put(19.5,34.5){\line(1,0){13.25}}
\put(41.25,43.25){\line(1,0){11.25}}
\put(19.25,43){\line(1,0){13.25}}
\put(38.75,35){\line(1,0){13.75}}
\put(61.25,43.25){\line(1,1){8.75}}
\put(70,52){\line(1,0){14.75}}
\put(18.5,52){\line(1,0){41}}
\multiput(59.5,52)(.033505155,-.043814433){97}{\line(0,-1){.043814433}}
\put(58.25,35.25){\line(1,0){33.75}}
\multiput(92,35.25)(.033505155,.038659794){97}{\line(0,1){.038659794}}
\multiput(64.5,45)(.03289474,-.04605263){38}{\line(0,-1){.04605263}}
\put(65.75,43.25){\line(1,0){19}}
\multiput(84.5,43.5)(.0346153846,.0336538462){260}{\line(1,0){.0346153846}}
\multiput(84.75,52)(.03370787,-.03651685){89}{\line(0,-1){.03651685}}
\multiput(52.5,43)(.033653846,-.046474359){156}{\line(0,-1){.046474359}}
\multiput(52.5,35)(.03353659,.03353659){82}{\line(0,1){.03353659}}
\multiput(56.75,39)(.035447761,.03358209){134}{\line(1,0){.035447761}}
\multiput(32.25,43)(.033602151,-.041666667){186}{\line(0,-1){.041666667}}
\multiput(32.75,34.75)(.03333333,.03333333){75}{\line(0,1){.03333333}}
\put(37,39){\line(1,1){4.25}}
\put(99.75,35.25){\line(1,0){45.75}}
\multiput(100,35.5)(-.0336990596,.0352664577){319}{\line(0,1){.0352664577}}
\multiput(97.25,41)(.0336363636,.04){275}{\line(0,1){.04}}
\put(106.5,52){\line(1,0){7.75}}
\put(121.25,44){\line(1,0){6.75}}
\put(128,44){\line(5,6){7.5}}
\put(135.5,53){\line(1,0){8.25}}
\put(93.25,52.25){\line(1,0){5.75}}
\multiput(99,52.25)(.03353659,-.04268293){82}{\line(0,-1){.04268293}}
\multiput(103,47)(.03333333,-.05){60}{\line(0,-1){.05}}
\put(105,44){\line(0,1){0}}
\put(105,44){\line(1,0){9.5}}
\multiput(114.5,44)(.033632287,.036995516){223}{\line(0,1){.036995516}}
\put(122,52.25){\line(1,0){5.25}}
\multiput(127.25,52.25)(.03353659,-.03963415){82}{\line(0,-1){.03963415}}
\multiput(131.5,47)(.03333333,-.04166667){60}{\line(0,-1){.04166667}}
\put(133.5,44.5){\line(1,0){10.75}}
\multiput(114.25,52.25)(.03370787,-.03651685){89}{\line(0,-1){.03651685}}
\multiput(121,44)(-.03333333,.04666667){75}{\line(0,1){.04666667}}
\label{3strand}
\end{picture}

\vspace{-2.5cm}

{\bf 6.} Examine properties of the mixing matrices (say, the eigenvalue hypothesis)
and the knot polynomials (say, various factorization properties,
differential expansions, recursions with the change of $R$ etc).

\bigskip

All this sounds simple, but is quite difficult in practice.
We comment on steps ${\bf 1-4}$ in section \ref{HV} and, more specifically, \ref{hv31},
provide some results of step ${\bf 5}$ in sections \ref{31pols}-\ref{evomn}
and describe some checks from step ${\bf 6}$ in section \ref{proppols}.
We end with a brief conclusion.

Throughout the paper we use the notation
\be
\{x\}\equiv x-{1\over x},\ \ \ \ \ \ D_n\equiv {\{Aq^n\}\over\{q\}},\ \ \ \ \ [n]\equiv {q^n-q^{-n}\over q-q^{-1}}
\ee
Let us also note that we use throughout the text the term "k-strand knot/link" which implies the knot/link whose braid representation with minimal number of strands is k-strand.

\section{Comments on the highest weight calculus \label{HV}}

One of the crucial decisions that makes calculations doable
is to extract the mixing matrices from the highest weights.
We already described the basis of this technique in \cite{MMMS21},
nowadays step {\bf 2} from the above list is provided by a fast working computer program,
which finds highest weight of $Y\in R_1\otimes R_2$, the same program
is used at step {\bf 4}.
We now briefly describe the new aspects, which were not clear
enough at the time of \cite{MMMS21}, other details can be found in that paper.

We distinguish between elementary and advanced levels of the method:
most calculations for $[3,1]$ were performed at the former level,
but further work on higher representations can be hard without going to the latter one.

\subsection{Elementary level}

\noindent

$\bullet$ Embed all representations in tensor powers $[1]^{\otimes M}$ of the fundamental representation
and parameterize elements of the Verma modules by number sequences.
For example, for the fundamental representation $M=1$, it has the highest weight $(0)$
and its other elements are $(1),(2),\ldots $. They are generated from $(0)$ by action of
the lowering Chevalley operators $\hat T_{-a}: \ (a-1)\longrightarrow (a)$.
Similarly, for the symmetric representation $[r]$ the highest weight is a sequence of $M=r$
zeroes $(00\ldots 0)$ and elements of the Verma module are $q$-symmetric linear combinations
$(10\ldots0)+q^{-1}(01\ldots 0)+\ldots + q^{1-r}(00\ldots 1)$ and so on.
The first antisymmetric representation $[1,1]$ has the highest vector $(10)-q(01)$.
The action of $\hat T_{-a}$ for $M>1$ is defined by comultiplication.

\bigskip

$\bullet$ Elements of the Verma modules are generated from the highest weights by action of the
Chevalley generators $\hat T_{-a}$ and their {\it ordered} products $\hat T_{-A} = \prod_i \hat T_{-a_i}$.
The sets of ordered sequences $A$  are {\it not} Young diagrams
(like they were in the case of Heisenberg or Virasoro Fock modules),
they can seem to be exponentially growing with the level $|A|=\sum_i a_i$,
but this is not true if one takes into account the conditions
\be
\hat T_{-a}\hat T_{-b}=\hat T_{-b}\hat T_{-a}, \ \ \ \ \ \forall \, a,b: \  |a-b|<2
\label{commfar}
\ee
and the Serre relations
\be
\hat T_{-a}^2\hat T_{-a-1}\ +\ \hat T_{-a-1}\hat T_{-a}^2=(q+q^{-1})\,\hat T_{-a}\hat T_{-a-1}\hat T_{-a},
\ \ \ \ \ \forall \,a
\label{Ser}
\ee
guaranteeing the PBW property for all simple algebras with allowed Dynkin diagrams.
In fact, for the $sl_N$ algebras the sets $A$ are closer to the $3d$ Young diagrams (plane partitions),
especially, if one does not restrict $N$: the Verma module for (continuous) $gl_\infty$ is naturally similar
to that of the double-affine DIM($gl_1$) \cite{FJMM}.
In practice, imposing (\ref{commfar}) and (\ref{Ser}) is indeed important for $R=[3,1]$
to make the problem solvable for finite time.
One can alternatively build the Verma modules with the help of all Borel generators, not Chevalley,
but we actually used the latter way.

\bigskip

$\bullet$ Highest weights are the vectors in $[1]^{\otimes M}$
annihilated by all rising operators $\hat T_a$, they
are parameterized by the Young diagrams of size $M$.

\bigskip

$\bullet$
The highest weight $V_R$ is a linear combination of sequences which contain
definite amounts of zeroes, units, twos etc.
These quantities are directly dictated by the Young diagram:
for $R=\{r_1\geq r_2\geq \ldots\}$ the number of zeroes is $r_1$,
the number of units is $r_2$ and so on.
For example, $V_{[5,1,1,1]}$ is a combination of sequences with
$\#(0)=5, \ \#(1)=1,\ \#(2)=1, \ \#(3)=1$.
We call collection of these entries the {\it type} $t_R$ of representation $R$,
e.g. $t_{[5,1,1,1]}=\{00000123\}$.
The highest weight $V_R$ is a certain linear combination of these sequences with different orderings.

\bigskip

$\bullet$ The highest weight $V_Y$ of representation $Y\in R_1\otimes R_2$ can be obtained
by acting with the lowering operators on the tensor product of the highest weights of $R_1$ and $R_2$:
\be
V_Y  \in  \oplus \Big( \hat T_{-A}V_{R_1}\otimes \hat T_{-B}V_{R_2}\Big),
\ee
The question is how to choose $A$ and $B$.

\bigskip

$\bullet$
The first criterion is simple:
one can look at the difference of {\it types} of $Y$ and $R_1$ and $R_2$.
For example, for $[5,1,1,1]\in [3,1]\otimes [3,1]$ one should compare
$t_{[5,1,1,1]}=\{00000123\}$ with $t_{[3,1]}\cup t_{[3,1]} = \{0001\}\cup\{0001\} =
\{00000011\}$.
To get the former from the latter, one should apply $\hat T_{-1}\hat T_{-2}\hat T_{-2}\hat T_{-3}$,
which means that together $A\cup B = \{1,2,2,3\}$.
Applying this criterion, one restricts the set of indices $\{a_i,b_i\}$
in the pair $(A,B)$, and (\ref{commfar})+(\ref{Ser}) reduce the ordering freedom.
Still, there are many terms of this type, differing by the positions of different
operators, and they can enter with arbitrary coefficients.

\bigskip

$\bullet$
To fix the coefficients, we apply the second criterion: the highest weight condition
\be
T_{a} V_Y = 0 \ \ \ \forall a
\label{hvcond}
\ee
If there are no multiplicities, i.e. $Y$ appears in $R_1\otimes R_2$ exactly once,
the coefficients are defined unambiguously modulo a common normalization factor.

\bigskip

$\bullet$
Find the highest weights of representations $Q \in Y\otimes R$ and $Q \in R\otimes Y$.
If $Q$ has multiplicities, then the highest weights can be chosen
in an arbitrary way.
This freedom affects the form of the Racah/mixing matrix, but not the answer for the knot polynomial.

\bigskip

$\bullet$ The mixing matrices are unitary if the highest weights are normalized.
Technically simplest is the norm
\be
\left|\left| \sum_{a_1,\ldots,a_M} c_{a_1\ldots a_M}(q)\cdot \{a_1,\ldots,a_M\}\ \right|\right|^2 =
\sum_{a_1,\ldots,a_M} \Big(c_{a_1\ldots a_M}(q)\Big)^2
\label{norm}
\ee

\bigskip

$\bullet$
The issue of multiplicities does not matter for the knot polynomials, because ${\cal R}$-matrix is
diagonal in representation $Y$, so the basis in the multi-dimensional
space of highest vectors can be chosen arbitrarily.
There is, however, one important exception: for $R_1=R_2=R$
one should distinguish between $Y_\pm$ belonging to $q$-symmetric and $q$-antisymmetric
squares of $R$, because the corresponding eigenvalues of ${\cal R}$
have different signs and do not coincide.
This is not a simple task, but there is a simple solution:
if one rotates the unitary mixing matrix ${\cal U}_Y$ into {\it symmetric}
(simply symmetric, with no reference to $q$-symmetry),
it automatically separates $Y_+$ from $Y_-$.

\subsection{Advanced level
\label{adval}}

The method of the previous subsection provides
highest weight vectors $V_Y$ as linear combinations of integer-number sequences from
$[1]^{\otimes M}$, which often contain enormously many terms (not exponentially growing,
but still too much).
The mixing matrices define linear dependencies between such combinations and they are difficult
to find even on powerful computers, despite it is just a linear algebra problem.

The following procedure helps to tame these linear combinations by
noting that they are made from certain standard pieces so that one can combine
substantially smaller number pieces
rather than the original number sequences:

\bigskip

$\bullet$ Parametrization by the number sequences can be converted into the one by polynomials of
$M$ auxiliary $x$-variables by the rule
$(a_1,a_2,\ldots a_M) \longleftrightarrow x_1^{a_1}x_2^{a_2}\ldots x_M^{a_M}$.

\bigskip

$\bullet$ The highest weights of totally antisymmetric representations $[1^r]$
are totally $q$-antisymmetric polynomials in $x$, i.e. certain quantum deformations
of the Vandermonde determinants $\Delta^{(r)}=\Delta_{1\ldots r} =\prod_{i<j}^r(x_i-x_j)$.
The point now is that the highest weights of arbitrary representations $R$
are expressed through the same determinants.

\bigskip

$\bullet$ The Young diagram $R =\{r_1\geq r_2\geq \ldots\}$ has $r_i$ as the heights of its columns.
The lengths of its lines are similar parameters of the transposed diagram
$R' = \{r_1'\geq r_2'\geq \ldots\}$.
These can be associated with totally antisymmetric representations $[1^{r_i'}]$, and
the highest weight $V_R(x)$ in $x$ representation is just a linear combination
of $\prod_i \Delta^{(r'_i)}$:
\be
V_R = \sum_{\sigma\in S_{|R|}} C_\sigma \cdot \Delta^{(r'_1)}_{ {\sigma_1}\ldots  {\sigma_{r'_1}}}
\cdot \Delta^{(r'_2)}_{ {\sigma_{r'_1+1}}\ldots  {\sigma_{r'_1+r'_2}}}\cdot \ldots
\ee
The sum goes over different distributions of $|R|$ variables $x_i$ between different
Vandermonde determinants, and the coefficients $C_\sigma$ are determined by the highest weight condition
(that the sum is annihilated by all raising operators $\hat T_a$).

\bigskip

$\bullet$
The Vandermonde decomposition is highly ambiguous, because there are many linear relations
between products of the Vandermonde determinants of a given type $R'$,
but this ambiguity does not change the highest weight itself.
Of course, in the case of non-trivial multiplicities, there are several
linear independent solutions to the highest weight condition.

\bigskip

$\bullet$ These expressions for the highest weights
can be straightforwardly quantized ($q$-deformed) keeping the highest weights.
We do not explain the quantization procedure in the present text: it is a separate
long story of its own interest.

\section{Specification to the case of $R=[3,1]$
\label{hv31}}

\subsection{Decomposition of the square $[3,1]^{\otimes 2}$ and highest weights}

The square of $[3,1]\otimes [3,1]$ contains $11$ different representations,
two of them twice, but we can distinguish between them, because they belong two
symmetric and antisymmetric squares, thus, for our purposes, there are $13$ different
representations, all with unit multiplicities (true multiplicities occur for the first
time in the square of $R=[4,2]$):
\be
\l[3,1]^{\otimes 2} =
[6,2]+[6,1,1]+[5,3]+2\cdot [5,2,1]+[5,1,1,1]+[4,4]+2\cdot[4,3,1]+[4,2,2]+[4,2,1,1]+[3,3,2]+[3,3,1,1]
\label{square31}
\ee
In what follows, we mark representations by indices $\pm$ depending on their belonging to symmetric
or antisymmetric squares.
The corresponding ${\cal R}$-matrix eigenvalues are:
{
\be
\lambda_{[6,2]}=q^{14}, \ \ \ \  \lambda_{[6,1,1]}=- {q^{12}}, \ \ \ \
\lambda_{[5,3]}=- {q^{10}}, \ \ \ \  \lambda_{[5,2,1]_\pm}=\pm {q^7}, \ \ \ \
\lambda_{[5,1,1,1]}= {q^4}, \ \ \ \
\lambda_{[4,4]}= {q^8},
\nn \\
\lambda_{[4,3,1]_\pm}=\pm  {q^4}, \ \ \ \
\lambda_{[4,2,2]}= {q^2}, \ \ \ \  \lambda_{[4,2,1,1]}=- {1}, \ \ \ \
\lambda_{[3,3,2]}=- {1}, \ \ \ \  \lambda_{[3,3,1,1]}= q^{-2}
\label{rev31}
\ee
}
They should all be additionally divided by $q^{4\varkappa_R} A^{|R|}=q^8A^4$ to provide the knot polynomials
in the topological framing.

If $[3,1]\otimes [3,1]$ is represented by $\Dq_{x_1,x_2}\otimes \Dq_{x_5,x_6} = \Dq_{12}\Dq_{56}$, then
the highest weights of the emerging representations are (we take them unreduced to simplify
both formulas and actual calculations: normalization of the intermediate representations $Y$ does
not affect the answers for mixing matrices):
\be
{\footnotesize
\begin{array}{rl}
V_{[6,2]_+} &= \Dq_{12}\Dq_{56} \nn \\  \nn \\
V_{[6,1,1]_-} &= \Dq_{125}-\Dq_{126} = \Dq_{256}-\Dq_{156} \nn \\  \nn\\
V_{[5,3]_-}& = \Dq_{12}\Big(\Dq_{37}\Dq_{56}  +\Dq_{47}\Dq_{56}+\Dq_{38}\Dq_{56} +\Dq_{48}\Dq_{56}\Big) \nn \\   \nn\\
V_{[5,2,1]_+}&= 4\Big(\Dq_{127}\Dq_{56}+\Dq_{128}\Dq_{56} + \Dq_{12}\Dq_{356}+\Dq_{12}\Dq_{456}\Big)
- 3\Big( \Dq_{123}\Dq_{56}+\Dq_{124}\Dq_{56}+ \Dq_{12}\Dq_{567}+\Dq_{12}\Dq_{568}\Big)
\nn \\      \nn\\
V_{[5,2,1]_-}&=4\Big( \Dq_{127}\Dq_{56}+\Dq_{128}\Dq_{56} -  \Dq_{12}\Dq_{356}-\Dq_{12}\Dq_{456}\Big)
-  \Big(\Dq_{123}\Dq_{56}+\Dq_{124}\Dq_{56}- \Dq_{12}\Dq_{567}-\Dq_{12}\Dq_{568} \Big)
\nn \\  \nn\\
V_{[5,1,1,1]_+} &= 4\Dq_{1256} +\Big( \Dq_{1235}+\Dq_{1245}- \Dq_{1236}-\Dq_{1246}\Big)
-\Big( \Dq_{1567}+\Dq_{1568}- \Dq_{2567}-\Dq_{2568}\Big) \nn \\   \nn\\
V_{[4,4]_-} &= \Dq_{12}\Big( \Dq_{38}\Dq_{47}\Dq_{56}+\Dq_{37}\Dq_{48}\Dq_{56}\Big) \nn \\  \nn\\
V_{[4,3,1]_+}  &=   \Dq_{12}\Dq_{356}\Dq_{47} -\Dq_{127}\Dq_{38}\Dq_{56}
+ \ (3\leftrightarrow 4) + (7\leftrightarrow 8)   \nn \\    \nn\\
V_{[4,3,1]_-} &=   \Dq_{12}\Dq_{356}\Dq_{47} +\Dq_{127}\Dq_{38}\Dq_{56} -\Dq_{125}\Dq_{67}\Dq_{38}
+\Dq_{126}\Dq_{57}\Dq_{38}
- \Dq_{125}\Dq_{36}\Dq_{47}+\Dq_{126}\Dq_{35}\Dq_{47}
+ \ (3\leftrightarrow 4) + (7\leftrightarrow 8)   \nn \\   \nn\\
V_{[4,2,2]_+} &=
-8\Big( \Dq_{127}\Dq_{356}+ \Dq_{127}\Dq_{456}+ \Dq_{128}\Dq_{356}+\Dq_{128}\Dq_{456}\Big)
 + \Big( \Dq_{123}\Dq_{567} +\Dq_{124}\Dq_{567}+ \Dq_{123}\Dq_{568} + \Dq_{124}\Dq_{568}\Big)
+\nn \\ & \ \ \ \ \ \ \ \ \ \ \
+4\Big(\Dq_{123}\Dq_{456}+ \Dq_{124}\Dq_{356}+ \Dq_{127}\Dq_{568}+ \Dq_{128}\Dq_{567}\Big)
 \nn \\  \nn\\
V_{[4,2,1,1]} &  = 2\Dq_{1256}\Dq_{37} - \Big(\Dq_{1237}\Dq_{56}+\Dq_{12}\Dq_{3567}\Big)
+ 2 \Big(\Dq_{1257}\Dq_{38}-\Dq_{1267}\Dq_{38}+\Dq_{1356}\Dq_{47}-\Dq_{2356}\Dq_{47}\Big)
+ \nn \\ &
 + \Big(\Dq_{1257}\Dq_{36}-\Dq_{1267}\Dq_{35}+\Dq_{1356}\Dq_{27}-\Dq_{2356}\Dq_{17}\Big)
+ \ (3\leftrightarrow 4) + (7\leftrightarrow 8)
\nn \\   \nn\\
V_{[3,3,2]}& = \Dq_{125}\Dq_{367}\Dq_{48}-\Dq_{126}\Dq_{357}\Dq_{48} - 3\cdot \Dq_{127}\Dq_{356}\Dq_{48}
+ \ (3\leftrightarrow 4) + (7\leftrightarrow 8)\nn \\ \nn\\
V_{[3,3,1,1]}& =3\Big( \Dq_{1237}\Dq_{48}\Dq_{56}+\Dq_{12}\Dq_{3567}\Dq_{48}\Big) - 6\Dq_{1256}\Dq_{37}\Dq_{48}
-\nn \\
& -5\Big(\Dq_{1257}\Dq_{36}\Dq_{48}-\Dq_{1267}\Dq_{35}\Dq_{48}
+\Dq_{1356}\Dq_{27}\Dq_{48}-\Dq_{2356}\Dq_{17}\Dq_{48}\Big)
+ \ (3\leftrightarrow 4) + (7\leftrightarrow 8)
\end{array}
}
\ee
The symmetrizations in $(3\leftrightarrow 4)$ and $ (7\leftrightarrow 8)$ are done independently,
i.e. each explicitly written term is substituted by the {\it four} ones.
As already mentioned, there are many different ways to represent the r.h.s.
because there are many linear relations between $\Delta$-products of a given type
and the choice of basis is somewhat arbitrary.
Also we do not provide here the exact definitions of $q$-deformed $\Dq$'s,
thus these formulas are mostly for demonstrative purposes.
They, however, can be directly used for $q=1$, i.e. for studying the {\it classical} Racah matrices, which is by itself quite a non-trivial problem for non-symmetric representations.

\subsection{Representations in the cube $[3,1]^{\otimes 3}$}

The representation content of the cube is now
\be
{\footnotesize
\begin{array}{l}
[3,1]\otimes\Big([3,1]\otimes[3,1]\Big) \ =
\\ \\
\!\!\!\!\!\!\!\!\!\!
=[3,1]\otimes\Big([6, 2]+[6, 1, 1]+[5, 3]+2\cdot[5, 2, 1]+[5, 1, 1, 1]+[4, 4]
+2\cdot[4, 3, 1]+[4, 2, 2]+[4, 2, 1, 1] + [3,3,2]+[3,3,1,1] \Big)=
\\  \\
=\Big([9, 3]+[9, 2, 1]+[8, 4]+[8, 3, 1]+[8, 3, 1]+[8, 2, 2]+[8, 2, 1, 1]+[7, 5]+2\cdot[7, 4, 1]+2\cdot[7, 3, 2]+[7, 2, 2, 1]+
\\
+[7, 3, 1, 1]+[6, 5, 1]+[6, 4, 2]+[6, 4, 1, 1]+[6, 3, 3]+[6, 3, 2, 1]\Big)+
\\  \\
+\Big([9, 2, 1]+[9, 1, 1, 1]+[8, 3, 1]+[8, 2, 2]+2\cdot[8, 2, 1, 1]+[8, 1, 1, 1, 1]+[7, 4, 1]+[7, 3, 2]+2\cdot[7, 3, 1, 1]+
\\
+[7, 2, 2, 1]+[7, 2, 1, 1, 1]+[6, 4, 2]+[6, 4, 1, 1]+[6, 3, 2, 1]+[6, 3, 1, 1, 1]\Big)+
\\  \\
+\Big([8, 4]+[8, 3, 1]+[7, 5]+2\cdot[7, 4, 1]+[7, 3, 2]+[7, 3, 1, 1]+[6, 6]+2\cdot[6, 5, 1]+2\cdot[6, 4, 2]+[6, 4, 1, 1]+[6, 3, 3]+
\\
+[6, 3, 2, 1]+[5, 5, 2]+[5, 5, 1, 1]+[5, 4, 3]+[5, 4, 2, 1]+[5, 3, 3, 1]\Big)+
\\ \\
+2\cdot\Big([8, 3, 1]+[8, 2, 2]+[8, 2, 1, 1]+[7, 4, 1]+2\cdot[7, 3, 2]+2\cdot[7, 3, 1, 1]+2\cdot[7, 2, 2, 1]+[7, 2, 1, 1, 1]+[6, 5, 1]+
\\
+2\cdot[6, 4, 2]+2\cdot[6, 4, 1, 1]+[6, 3, 3]+3\cdot[6, 3, 2, 1]+[6, 3, 1, 1, 1]+[6, 2, 2, 2]+[6, 2, 2, 1, 1]+[5, 5, 2]+[5, 5, 1, 1]+
\\
+[5, 4, 3]+2\cdot[5, 4, 2, 1]+[5, 4, 1, 1, 1]+[5, 3, 2, 2]+[5, 3, 3, 1]+[5, 3, 2, 1, 1]\Big)+
\\  \\
+\Big([8, 2, 1, 1]+[8, 1, 1, 1, 1]+[7, 3, 1, 1]+[7, 2, 2, 1]+2\cdot[7, 2, 1, 1, 1]+[7, 1, 1, 1, 1, 1]+[6, 4, 1, 1]+[6, 3, 2, 1]+
\\
+2\cdot[6, 3, 1, 1, 1]+[6, 2, 2, 1, 1]+[6, 2, 1, 1, 1, 1]+[5, 4, 2, 1]+[5, 4, 1, 1, 1]+[5, 3, 2, 1, 1]+[5, 3, 1, 1, 1, 1]\Big)+
\\   \\
+\Big([7, 5]+[7, 4, 1]+[6, 5, 1]+[6, 4, 2]+[6, 4, 1, 1]+[5, 5, 2]+[5, 4, 3]+[5, 4, 2, 1]+[4, 4, 3, 1]\Big)+
\\  \\
+2\cdot\Big([7, 4, 1]+[7, 3, 2]+[7, 3, 1, 1]+[6, 5, 1]+2\cdot[6, 4, 2]+2\cdot[6, 4, 1, 1]+[6, 3, 3]+2\cdot[6, 3, 2, 1]+[6, 3, 1, 1, 1]+
\\
+[5, 5, 2]+[5, 5, 1, 1]+2\cdot[5, 4, 3]+3\cdot[5, 4, 2, 1]+[5, 4, 1, 1, 1]+[5, 3, 2, 2]+2\cdot[5, 3, 3, 1]+[5, 3, 2, 1, 1]+
\\
+[4, 4, 4]+[4, 4, 2, 2]+2\cdot[4, 4, 3, 1]+[4, 4, 2, 1, 1]+[4, 3, 3, 2]+[4, 3, 3, 1, 1]\Big)+
\\    \\
+\Big([7, 3, 2]+[7, 2, 2, 1]+[6, 4, 2]+[6, 3, 3]+2\cdot[6, 3, 2, 1]+[6, 2, 2, 2]+[6, 2, 2, 1, 1]+[5, 5, 2]+[5, 4, 3]+2\cdot[5, 4, 2, 1]+
\\
+2\cdot[5, 3, 2, 2]+[5, 3, 3, 1]+[5, 3, 2, 1, 1]+[5, 2, 2, 2, 1]+[4, 4, 2, 2]+[4, 4, 3, 1]+[4, 4, 2, 1, 1]+[4, 3, 3, 2]+[4, 3, 2, 2, 1]\Big)+
\\ \\
+\Big([7, 3, 1, 1]+[7, 2, 2, 1]+[7, 2, 1, 1, 1]+[6, 4, 1, 1]+2\cdot[6, 3, 2, 1]+2\cdot[6, 3, 1, 1, 1]+[6, 2, 2, 2]+2\cdot[6, 2, 2, 1, 1]+
\\
+[6, 2, 1, 1, 1, 1]+[5, 5, 1, 1]+2\cdot[5, 4, 2, 1]+2\cdot[5, 4, 1, 1, 1]+[5, 3, 3, 1]+[5, 3, 2, 2]+3\cdot[5, 3, 2, 1, 1]+[5, 3, 1, 1, 1, 1]+
\\
+[5, 2, 2, 2, 1]+[5, 2, 2, 1, 1, 1]+[4, 4, 3, 1]+[4, 4, 2, 2]+2\cdot[4, 4, 2, 1, 1]+[4, 4, 1, 1, 1, 1]+[4, 3, 3, 1, 1]+
\phantom{Big|^{5^5}}\\
+[4, 3, 2, 2, 1]+[4, 3, 2, 1, 1, 1]\Big)+
\\ \\
+\Big([6, 4, 2]+[6, 3, 3]+[6, 3, 2, 1]+[5, 4, 3]+[5, 4, 2, 1]+2\cdot[5, 3, 3, 1]+[5, 3, 2, 2]+[5, 3, 2, 1, 1]+[4, 4, 3, 1]+[4, 4, 2, 2]+
\\
+2\cdot[4, 3, 3, 2]+[4, 3, 3, 1, 1]+[4, 3, 2, 2, 1]+[3, 3, 3, 3]+[3, 3, 3, 2, 1]\Big)+
\\  \\
+\Big([6, 4, 1, 1]+[6, 3, 2, 1]+[6, 3, 1, 1, 1]+[5, 4, 2, 1]+[5, 4, 1, 1, 1]+[5, 3, 3, 1]+[5, 3, 2, 2]+2\cdot[5, 3, 2, 1, 1]+[5, 3, 1, 1, 1, 1]+
\\
+[4, 4, 3, 1]+[4, 4, 2, 1, 1]+[4, 3, 3, 2]+2\cdot[4, 3, 3, 1, 1]+[4, 3, 2, 2, 1]+[4, 3, 2, 1, 1, 1]+[3, 3, 3, 2, 1]+[3, 3, 3, 1, 1, 1]\Big)
\end{array}
}
\label{decocube}
\ee
and the highest weights are too numerous to be listed here, see \cite{knotebook}.

Each block of lines here is associated with the corresponding intermediate representation:
$[6,2], \ [6,1,1], \ \ldots$ from (\ref{square31}),
and most representations appear in several blocks in the table below.
The number of times the representation $Q$ enters (\ref{decocube}) is given in the first column of the table,
and it is the size of the mixing matrix ${\cal U}_Q$ which we need to calculate.
Clearly, there are quite a few ($40$) matrices of non-unit size, some are quite big.
Only $26$ of them (up to size $6$) can be found from the eigenvalue hypothesis of
\cite{IMMMev} and its recent generalization in \cite{univev},
all the rest had to be calculated by the methods of sec.2 (actually we did so
also for the matrices of sizes $5$ and $6$ to double check the eigenvalue hypothesis).

{\footnotesize
\be
\begin{array}{|c|c|c|}
\hline
&&\text{number of}\\
\text{matrix size} & Q &  \\
&&\text{matrices}\\ \hline && \\
1 & [9{,} 3], \  [9{,} 1{,} 1{,} 1], \  [7{,} 1{,} 1{,} 1{,} 1{,} 1], \  [6{,} 6], \  [5{,} 2{,} 2{,} 1{,} 1{,} 1], \  [4{,} 4{,} 1{,} 1{,} 1{,} 1], \  [3{,} 3{,} 3{,} 3], \  [3{,} 3{,} 3{,} 1{,} 1{,} 1] & 8 \\
&&\\ \hline && \\
2 & [9{,} 2{,} 1], \  [8{,} 4], \  [8{,} 1{,} 1{,} 1{,} 1], \  [6{,} 2{,} 1{,} 1{,} 1{,} 1], \  [5{,} 2{,} 2{,} 2{,} 1], \  [4{,} 4{,} 4], \  [4{,} 3{,} 2{,} 1{,} 1{,} 1], \  [3{,} 3{,} 3{,} 2{,} 1] & 8 \\
&&\\ \hline && \\
3 & [7{,} 5], \  [5{,} 3{,} 1{,} 1{,} 1{,} 1] & 2 \\
&&\\ \hline && \\
4 & [8{,} 2{,} 2], \  [6{,} 2{,} 2{,} 2], \  [4{,} 3{,} 2{,} 2{,} 1] & 3 \\
&&\\ \hline && \\
5 & [4{,} 4{,} 2{,} 2] & 1 \\
&&\\ \hline && \\
6 & [8{,} 3{,} 1], \  [8{,} 2{,} 1{,} 1], \  [7{,} 2{,} 1{,} 1{,} 1], \  [6{,} 2{,} 2{,} 1{,} 1], \  [5{,} 5{,} 1{,} 1], \  [4{,} 4{,} 2{,} 1{,} 1], \  [4{,} 3{,} 3{,} 2], \  [4{,} 3{,} 3{,} 1{,} 1] & 8 \\
&&\\ \hline && \\
7 & [5{,} 5{,} 2] & 1 \\
&&\\ \hline && \\
8 & [6{,} 5{,} 1], \  [6{,} 3{,} 3], \  [5{,} 4{,} 1{,} 1{,} 1] & 3 \\
&&\\ \hline && \\
9 & [7{,} 2{,} 2{,} 1], \  [5{,} 3{,} 2{,} 2], \  [4{,} 4{,} 3{,} 1] & 3 \\
&&\\ \hline && \\
10 & [7{,} 4{,} 1], \  [6{,} 3{,} 1{,} 1{,} 1], \  [5{,} 4{,} 3] & 3 \\
&&\\ \hline && \\
11 & [7{,} 3{,} 2] & 1 \\
&&\\ \hline && \\
12 & [7{,} 3{,} 1{,} 1], \  [5{,} 3{,} 3{,} 1], \  [5{,} 3{,} 2{,} 1{,} 1] & 3 \\
&&\\ \hline && \\
15 & [6{,} 4{,} 2], \  [6{,} 4{,} 1{,} 1] & 2 \\
&&\\ \hline && \\
19 & [5{,} 4{,} 2{,} 1] & 1 \\
&&\\ \hline && \\
20 & [6{,}3{,}2{,}1] & 1 \\
&&\\ \hline
 \end{array}
\ee
}

\section{Knot polynomials in representation $[3,1]$ \label{31pols}}

\subsection{2-strand (torus) knots and links
\label{2strand}}

Among the knots and links having 3-strand braid representation there are many composites consisting of 2-strand
torus components.
The reduced HOMFLY polynomials for the composites are just products of those
for the constituents and, for the sake of completeness,
we complement (\ref{3strafla}) by its simple $2$-strand counterpart:
\be
H_R^{(n)} = \sum_{Y\in R\otimes R} \frac{d_Y}{d_R} \cdot
\left(\frac{\epsilon_Y q^{\varkappa_Y}}{q^{4\varkappa_R}A^{|R|}}\right)^n
\label{2straH}
\ee
with Casimir eigenvalue $\varkappa_Y = \sum_{(i,j)\in Y} \big(i-j\big)$,
quantum dimension
\be
d_Y = {\rm dim}_q(Y)={\rm Schur}_Y\left(p_k=\{A^k\}/\{q^k\}\right)=
\prod_{(i,j)\in Y} \frac{\{Aq^{i-j}\}}{\{q^{1+{\rm arm}(i,j)+{\rm leg}(i,j)}\}},
\ee
and $\epsilon_Y=\pm 1$ depending on whether $Y$ belongs to the symmetric or antisymmetric
square of $R$.

\subsection{Abundance of 3-strand prime knots in Rolfsen table}

As to the {\it prime} knots given by 3-strand braids, their place in the entire set is clear from
the following table, where the minimal number of strands is shown for up to 10 crossings.
Explicitly listed are the 3-strand braid representations of knots,
and just the number of strands is given for everything else.
The non-arborescent knots are boldfaced.

\vspace{5mm}
{\footnotesize
\hspace{-12mm}
$
\begin{array}{|c|c|  }
\hline
&\\
{\rm knot}  & \text{braid index}
\\
&\\ \hline &\\
3_1  &   2 \\
&\\ \hline &\\
4_1 &  (1,-1,1,-1) \\
&\\ \hline &\\
5_1 &  2 \\
5_2 &  (3,1,-1,1) \\
&\\ \hline &\\
6_1 & 4  \\
6_2 &   (3,-1,1,-1) \\
6_3 &   ( 2,-1,1,-2) \\
&\\ \hline  &\\
7_1 & 2 \\
{7_2} &4\\
7_3 &    (5,1,-1,1) \\
{ 7_4} &4\\
7_5 &   (4,1,-1,2) \\
{7_6} &4\\
{7_7} &4\\
&\\ \hline  &\\
{8_1} &5\\
{8_2} &5,-1,1,-1\\
{8_3} &5\\
 {8_4} &4\\
{ 8_5} & 3,-1,3,-1\\
 {8_6} &4\\
 { 8_7} &4,-1,1,-2\\
{8_8} &4\\
 { 8_9} & 3,-1,1,-3\\
{ 8_{10}} & 3,-1,2,-2\\
{8_{11}} &4\\
{8_{12}} &5\\
{8_{13}} &4\\
{8_{14}} &4\\
{8_{15}} &4\\
 { 8_{16}} & 2,-1,2,-1,1,-1\\
  { 8_{17}} & 2,-1,1,-1,1,-2 \\
\bf { 8_{18}} & (1,-1)^4 \\
{ 8_{19}} & 3,1,3,1 \\
{ 8_{20}} &  3,-1,-3,-1\\
{ 8_{21}} & 3,1,-2,2 \\
\hline
\end{array}
$
\quad
$
\begin{array}{|c|c|  }
\hline
{9_1} & 2 \\
{9_2} &5\\
{9_3} &7,1,-1,1\\
{9_4} &4\\
{9_5} &5\\
{9_6} & 6,1,-1,2 \\
{9_7} &4\\
{9_8} &5\\
{9_9} & 5,1,-1,3 \\
{9_{10}} &4\\
{9_{11}} &4\\
{9_{12}} &5\\
{9_{13}} &4\\
{9_{14}} &5\\
{9_{15}} &5\\
{ 9_{16}} &4,2,-1,3\\
{9_{17}} &4\\
{9_{18}} &4\\
{9_{19}} &5\\
{9_{20}} &4\\
{9_{21}} &5\\
{ 9_{22}} &4\\
{9_{23}} &4\\
{ 9_{24}} &4\\
 { 9_{25}} &5\\
 {9_{26}} &4\\
 {9_{27}} &4\\
 { 9_{28}} &4\\
{ 9_{29}} &4\\
{ 9_{30}} &4\\
{9_{31}} &4\\
{ 9_{32}} &4\\
{ 9_{33}} &4\\
\bf { 9_{34}} &4\\
{ 9_{35}} &5\\
{ 9_{36}} &4\\
{ 9_{37}} &5\\
{ 9_{38}} &4\\
\bf { 9_{39}} &5\\
\bf { 9_{40}} &4\\
\bf { 9_{41}} &5\\
{ 9_{42}} &4\\
{ 9_{43}} &4\\
{9_{44}} &4\\
{ 9_{45}} &4\\
{ 9_{46}} &4\\
\bf { 9_{47}} &4\\
{9_{48}} &4\\
\bf {9_{49}} &4\\
\hline
\end{array}
$
\quad
$
\begin{array}{ |c|c|  }
\hline
{10_1} &6\\
{10_2}  &7,-1,1,-1 \\
{10_3} &6\\
{ 10_4} &5\\
{ 10_5} &6,-1,1,-2 \\
{ 10_6} &4\\
{ 10_7} &5\\
{ 10_8} &4\\
{ 10_9} &5,-1,1,-3 \\
{ 10_{10}} &5\\
{ 10_{11}} &5\\
{ 10_{12}} &4\\
{ 10_{13}} &6\\
{ 10_{14}} &4\\
{ 10_{15}} &4\\
{ 10_{16}} &5\\
{ 10_{17}} &4,-1,1,-4 \\
{ 10_{18}} &5\\
{ 10_{19}} &4\\
 { 10_{20}} &5\\
 { 10_{21}} &4\\
 { 10_{22}} &4\\
 { 10_{23}} &4\\
 { 10_{24}} &5\\
 { 10_{25}} &4\\
 { 10_{26}} &4\\
 { 10_{27}} &4\\
 { 10_{28}} &5\\
 { 10_{29}} &5\\
 { 10_{30}} &5\\
 { 10_{31}} &5\\
 { 10_{32}} &4\\
 { 10_{33}} &5\\
 { 10_{34}} &5\\
 { 10_{35}} &6\\
 { 10_{36}} &5\\
 { 10_{37}} &5\\
 { 10_{38}} &5\\
 { 10_{39}} &4\\
 { 10_{40}} &4\\
 { 10_{41}} &5\\
 { 10_{42}} &5\\
 { 10_{43}} &5\\
 { 10_{44}} &5\\
 { 10_{45}} &5\\
 { 10_{46}} &5,-1,3,-1 \\
 { 10_{47}} &5,-1,2,-2 \\
  { 10_{48}} &4,-2,1,-3 \\
  {10_{49}} &4\\
  { 10_{50}} &4\\
  { 10_{51}} &4\\
  { 10_{52}} &4\\
  { 10_{53}} &5\\
  { 10_{54}} &4\\
  { 10_{55}} &5\\
\hline
  \end{array}
  $
  \quad
  $
  \begin{array}{|c|c|}
  \hline
  { 10_{56}} &4\\
  { 10_{57}} &4\\
  { 10_{58}} &6\\
  { 10_{59}} &5\\
  { 10_{60}} &5\\
  { 10_{61}} &4\\
  { 10_{62}} &4,-1,3,-2 \\
  { 10_{63}} &5\\
  { 10_{64}} &3,-1,3,-3  \\
  { 10_{65}} &4\\
  { 10_{66}} &4\\
  { 10_{67}} &5\\
  { 10_{68}} &5\\
  { 10_{69}} &5\\
  { 10_{70}} &5\\
{ 10_{71}} &5\\
{ 10_{72}} &4\\
{ 10_{73}} &5\\
{ 10_{74}} &5\\
{ 10_{75}} &5\\
{ 10_{76}} &4\\
{ 10_{77}} &4\\
{ 10_{78}} &5\\
10_{79} &3,-2,2,-3 \\
{ 10_{80}} &4\\
{ 10_{81}} &5\\
10_{82} &4,-1,1,-1,1,-2\\
{ 10_{83}} &4\\
{ 10_{84}} &4\\
10_{85} &4,-1,2,-1,1,-1\\
{ 10_{86}} &4\\
{ 10_{87}} &4\\
{ 10_{88}} &5\\
{ 10_{89}} &5\\
{ 10_{90}} &4\\
10_{91} &3,-1,1,-2,1,-2 \\
{ 10_{92}} &4\\
{ 10_{93}} &4\\
10_{94} &3,-1,2,-2,1,-1\\
{ 10_{95}} &4\\
{ 10_{96}} &5\\
{ 10_{97}} &5\\
{ 10_{98}} &4\\
10_{99} &2,-1,2,-2,1,-2\\
\bf 10_{100}&3,-1,2,-1,2,-1\\
\bf { 10_{101}} &5\\
\bf { 10_{102}} &4\\
\bf { 10_{103}} &4\\
\bf 10_{104}&3,-2,1,-1,1,-2\\
\bf { 10_{105}} &5\\
\bf 10_{106}&3,-1,1,-1,2,-2\\
\bf { 10_{107}} &5\\
\bf { 10_{108}} &4\\
\bf 10_{109}&2,-1,1,-2,2,-2\\
\bf { 10_{110}} &5\\
\hline
\end{array}
  $
\quad
  $
  \begin{array}{|c|c|}
  \hline
\bf { 10_{111}} &4\\
\bf 10_{112}&3,-1,1,-1,1,-1,1,-1\\
\bf { 10_{113}} &4\\
\bf { 10_{114}} &4\\
\bf { 10_{115}} &5\\
\bf 10_{116}&2,-1,2,-1,1,-1,1,-1\\
\bf { 10_{117}} &4\\
\bf { 10_{118}} &2,-1,1,-1,1,-2,1,-1\\
\bf { 10_{119}} &4\\
\bf { 10_{120}} &5\\
\bf { 10_{121}} &4\\
\bf { 10_{122}} &4\\
\bf 10_{123}&(1,-1)^5\\
{10_{124}} &5,1,3,1 \\
{10_{125}} & 5,-1,-3,-1 \\
{10_{126}} & 5,1,-3,1\\
{10_{127}} &5,1,-2,2\\
{ 10_{128}} &4\\
{10_{129}} &4\\
{ 10_{130}} &4\\
{ 10_{131}} &4\\
{ 10_{132}} &4\\
{ 10_{133}} &4\\
{ 10_{134}} &4\\
{ 10_{135}} &4\\
{ 10_{136}} &4\\
{ 10_{137}} &5\\
{ 10_{138}} &5\\
{10_{139}} &4,1,3,2 \\
{ 10_{140}} &4\\
{10_{141}} &4,-1,-3,-2 \\
{ 10_{142}} &4\\
{ 10_{143}} &4,1,-3,2 \\
{ 10_{144}} &4\\
{ 10_{145}} &4\\
{ 10_{146}} &4\\
{ 10_{147}} &4\\
10_{148}&4,1,-2,1,-1,1 \\
10_{149}&4,1,-1,1,-1,2 \\
{ 10_{150}} &4\\
{ 10_{151}} &4\\
10_{152}&3,2,2,3 \\
{ 10_{153}} &4\\
{ 10_{154}} &4\\
\bf 10_{155}&3,1,-2,1,-2,1 \\
\bf { 10_{156}} &4\\
\bf 10_{157}&3,2,-1,1,-1,2 \\
\bf { 10_{158}} &4\\
\bf 10_{159}&3,1,-1,1,-2,2 \\
\bf { 10_{160}} &4\\
\bf 10_{161}&3,1,-1,1,2,2 \\
\bf { 10_{162}} &4\\
\bf { 10_{163}} &4\\
\bf { 10_{164}} &4\\
\bf { 10_{165}} &4\\
\hline
\end{array}
$
}

\subsection{Examples
\label{exaknots}}

Explicit expressions for the knot polynomials are quite lengthy,
we give just three examples in the two simplest
and one more complicated case:

\bigskip

{\bf Trefoil knot $3_1\,=\,(m_1=1,n_1=1,m_2=1,n_2=1)$ also known as the torus knot $T[3,2]$:}

\bigskip

{\footnotesize
$H_{[3,1]}^{3_1}={q}^{12}{A}^{16}+{ { \Big( -{q}^{30}-{q}^{28}-{q}^{26}-{q}^{24}-{q}^{20}-{q}^{18}-{q}^{16}-{q}^{14} \Big) {A}^{14}}{{q}^{-12}}}+ \Big( {q}^{34}+2\,{q}^{30}+2\,{q}^{28}+{q}^{26}+3\,{q}^{24}+2\,{q}^{22}+2\,{q}^{20}+4\,{q}^{18}+{q}^{16}+2\,{q}^{14}+\\{q}^{12}+{q}^{10}+{q}^{8}+{q}^{6} \Big) {A}^{12}{{q}^{-12}}+{{ \Big( -{q}^{34}-{q}^{32}-{q}^{30}-3\,{q}^{28}-{q}^{26}-3\,{q}^{24}-4\,{q}^{22}-{q}^{20}-4\,{q}^{18}-2\,{q}^{16}-{q}^{14}-4\,{q}^{12}-2\,{q}^{10}-{q}^{8}-2\,{q}^{6}-1 \Big) {A}^{10}}{{q}^{-12}}}+{{ \Big( {q}^{32}+2\,{q}^{28}+{q}^{26}+2\,{q}^{22}+{q}^{20}+{q}^{18}+2\,{q}^{16}+2\,{q}^{12}+{q}^{10}+{q}^{6}+{q}^{4}+1 \Big) {A}^{8}}{{q}^{-12}}}
$
}

\bigskip

{\bf Figure-eight knot $4_1\,=\,(m_1=1,n_1-1,m_2=1,n_2=-1)$:}

\bigskip

{\footnotesize
$H_{[3,1]}^{4_1}={q}^{8}{A}^{8}+ { \Big( -{q}^{32}-{q}^{28}-{q}^{26}+{q}^{24}-{q}^{22}-{q}^{16} \Big) {A}^{6}}{{q}^{-18}}+  \Big( {q}^{36}-{q}^{34}+3\,{q}^{32}+{q}^{30}-2\,{q}^{28}+4\,{q}^{26}-2\,{q}^{24}-{q}^{22}+5\,{q}^{20}-{q}^{18}+{q}^{16}+\\2\,{q}^{14}-{q}^{12}+{q}^{8} \Big) {A}^{4}{{q}^{-18}}+ { \Big( -2\,{q}^{36}-5\,{q}^{30}+5\,{q}^{28}-{q}^{26}-8\,{q}^{24}+7\,{q}^{22}-6\,{q}^{20}-5\,{q}^{18}+9\,{q}^{16}-6\,{q}^{14}-2\,{q}^{12}+3\,{q}^{10}-5\,{q}^{8}+{q}^{4}-{q}^{2} \Big) {A}^{2}}{{q}^{-18}}+ \Big({q}^{36}+2\,{q}^{34}-3\,{q}^{32}+4\,{q}^{30}+2\,{q}^{28}-7\,{q}^{26}+12\,{q}^{24}-{q}^{22}-8\,{q}^{20}+15\,{q}^{18}-8\,{q}^{16}-{q}^{14}+12\,{q}^{12}-7\,{q}^{10}+2\,{q}^{8}+4\,{q}^{6}-3\,{q}^{4}+2\,{q}^{2}+1\Big){{q}^{-18}}+ {-{q}^{34}+{q}^{32}-5\,{q}^{28}+3\,{q}^{26}-2\,{q}^{24}-6\,{q}^{22}+9\,{q}^{20}-5\,{q}^{18}-6\,{q}^{16}+7\,{q}^{14}-8\,{q}^{12}-{q}^{10}+5\,{q}^{8}-5\,{q}^{6}-2}{{q}^{-18}{A}^{-2}}  +  \Big({q}^{28}-{q}^{24}+2\,{q}^{22}+{q}^{20}-{q}^{18}+5\,{q}^{16}-{q}^{14}-2\,{q}^{12}+4\,{q}^{10}-2\,{q}^{8}+{q}^{6}+3\,{q}^{4}-{q}^{2}+1\Big){{q}^{-18}{A}^{-4}}+ \Big(-{q}^{20}-{q}^{14}+{q}^{12}-{q}^{10}-{q}^{8}-{q}^{4}\Big){{q}^{-18}{A}^{-6}}+ {{q}^{-8}{A}^{-8}}
$
}

\bigskip

{\bf Knot $10_{161}$ (non-arborescent, thick, {\rm but with relatively short $H_{[31]}$}):}

\bigskip

{\footnotesize
$H_{[3,1]}^{10_{161}}=\Big( {q}^{50}-{q}^{48}-{q}^{46}+3\,{q}^{44}-4\,{q}^{42}+2\,{q}^{40}+5\,{q}^{38}-9\,{q}^{36}+5\,{q}^{34}+4\,{q}^{32}-9\,{q}^{30}+9\,{q}^{28}-2\,{q}^{26}-6\,{q}^{24}+6\,{q}^{22}-2\,{q}^{20}+{q}^{16}-{q}^{14} \Big) {A}^{40}+ \Big( 2\,{q}^{52}-{q}^{50}+3\,{q}^{46}-4\,{q}^{44}+2\,{q}^{42}+4\,{q}^{40}-10\,{q}^{38}+8\,{q}^{36}+{q}^{34}-14\,{q}^{32}+16\,{q}^{30}-8\,{q}^{28}-6\,{q}^{26}+16\,{q}^{24}-11\,{q}^{22}+6\,{q}^{18}-5\,{q}^{16}+2\,{q}^{14}+3\,{q}^{12}-{q}^{10}-{q}^{8}+{q}^{6}+1 \Big) {A}^{38}+ \Big( -1-{q}^{62}+2\,{q}^{54}+6\,{q}^{48}-9\,{q}^{46}+9\,{q}^{38}-7\,{q}^{34}+8\,{q}^{42}-{q}^{58}-5\,{q}^{22}+2\,{q}^{20}+{q}^{26}+2\,{q}^{44}-3\,{q}^{50}-13\,{q}^{40}+{q}^{-2}-7\,{q}^{16}-4\,{q}^{52}+3\,{q}^{36}+6\,{q}^{24}+7\,{q}^{32}-{q}^{-10}+{q}^{12}-2\,{q}^{56}+3\,{q}^{8}+3\,{q}^{30}-{q}^{2}-2\,{q}^{6}+3\,{q}^{18}-5\,{q}^{28}-2\,{q}^{10}+{q}^{14} \Big) {A}^{36}+ \Big( 7+4\,{q}^{54}+2\,{q}^{48}-2\,{q}^{46}-7\,{q}^{38}+3\,{q}^{34}+2\,{q}^{60}-5\,{q}^{22}-13\,{q}^{20}-21\,{q}^{26}+3\,{q}^{-6}+{q}^{50}+2\,{q}^{40}+{q}^{66}-9\,{q}^{-2}-7\,{q}^{16}-2\,{q}^{52}-8\,{q}^{4}+15\,{q}^{24}-16\,{q}^{32}-2\,{q}^{-14}+{q}^{-10}+9\,{q}^{12}-7\,{q}^{8}+6\,{q}^{30}+{q}^{-4}-3\,{q}^{2}+12\,{q}^{6}+16\,{q}^{18}-4\,{q}^{-8}+3\,{q}^{28}-6\,{q}^{10}-8\,{q}^{14} \Big) {A}^{34}+ \Big( -8+{q}^{-26}-{q}^{64}+{q}^{62}-{q}^{54}-4\,{q}^{48}+3\,{q}^{46}-7\,{q}^{38}+6\,{q}^{34}-8\,{q}^{42}-14\,{q}^{22}+13\,{q}^{20}+{q}^{26}+{q}^{44}-7\,{q}^{-6}+4\,{q}^{50}+12\,{q}^{40}+9\,{q}^{-2}-7\,{q}^{16}-{q}^{52}+5\,{q}^{36}+10\,{q}^{4}+10\,{q}^{24}-10\,{q}^{32}+4\,{q}^{-14}-5\,{q}^{-12}+4\,{q}^{-10}-15\,{q}^{12}+{q}^{56}+4\,{q}^{8}+13\,{q}^{30}+2\,{q}^{-16}+2\,{q}^{2}-14\,{q}^{6}-4\,{q}^{18}+4\,{q}^{-8}-6\,{q}^{28}+{q}^{-22}+2\,{q}^{-20}-3\,{q}^{-18}+7\,{q}^{10}+14\,{q}^{14} \Big) {A}^{32}+ \Big( 5+{q}^{-30}+{q}^{64}-5\,{q}^{54}-2\,{q}^{48}+8\,{q}^{46}-4\,{q}^{34}-{q}^{-32}+4\,{q}^{42}-3\,{q}^{60}+2\,{q}^{58}-4\,{q}^{22}+28\,{q}^{20}+25\,{q}^{26}-4\,{q}^{44}+2\,{q}^{-6}-4\,{q}^{50}-{q}^{40}-{q}^{66}+3\,{q}^{-2}+5\,{q}^{16}+6\,{q}^{52}+7\,{q}^{36}+3\,{q}^{4}-9\,{q}^{24}+12\,{q}^{32}+7\,{q}^{-14}+{q}^{-12}-2\,{q}^{-10}-7\,{q}^{12}+{q}^{56}+9\,{q}^{8}+3\,{q}^{-24}+2\,{q}^{30}+6\,{q}^{-4}-3\,{q}^{-16}+4\,{q}^{2}+2\,{q}^{6}-15\,{q}^{18}+4\,{q}^{-8}-13\,{q}^{28}-4\,{q}^{-22}+{q}^{-20}-{q}^{-28}+3\,{q}^{-18}+7\,{q}^{10}+17\,{q}^{14} \Big) {A}^{30}+ \Big( -31-{q}^{-30}-3\,{q}^{-26}+{q}^{54}+{q}^{48}-5\,{q}^{46}+20\,{q}^{38}-2\,{q}^{34}-4\,{q}^{42}+{q}^{60}+26\,{q}^{22}+22\,{q}^{20}+30\,{q}^{26}+9\,{q}^{44}-20\,{q}^{-6}+3\,{q}^{50}-6\,{q}^{40}+7\,{q}^{-2}+34\,{q}^{16}-22\,{q}^{36}+9\,{q}^{4}+{q}^{-34}-55\,{q}^{24}+28\,{q}^{32}-3\,{q}^{-14}-8\,{q}^{-12}+2\,{q}^{-10}-40\,{q}^{12}+11\,{q}^{8}-{q}^{-24}-32\,{q}^{30}+8\,{q}^{-4}+15\,{q}^{2}-34\,{q}^{6}-53\,{q}^{18}-{q}^{-36}+4\,{q}^{-8}+8\,{q}^{28}+2\,{q}^{-22}-3\,{q}^{-20}+2\,{q}^{-28}-{q}^{-18}+21\,{q}^{10}+6\,{q}^{14} \Big) {A}^{28}+ \Big( 13-{q}^{-30}+{q}^{-26}-3\,{q}^{54}-7\,{q}^{48}+3\,{q}^{46}-11\,{q}^{38}+4\,{q}^{34}+{q}^{-32}-8\,{q}^{42}-{q}^{60}-{q}^{58}+21\,{q}^{22}-47\,{q}^{20}-25\,{q}^{26}-6\,{q}^{44}+12\,{q}^{-6}+5\,{q}^{40}+8\,{q}^{-2}+17\,{q}^{16}-2\,{q}^{52}-3\,{q}^{36}+8\,{q}^{4}-{q}^{-34}-{q}^{24}-16\,{q}^{32}-4\,{q}^{-14}+9\,{q}^{-12}-10\,{q}^{-10}+24\,{q}^{12}-36\,{q}^{8}-2\,{q}^{-24}-7\,{q}^{30}-23\,{q}^{-4}-{q}^{-16}-28\,{q}^{2}+18\,{q}^{6}+17\,{q}^{18}+17\,{q}^{28}+2\,{q}^{-22}-2\,{q}^{-20}-{q}^{-28}+5\,{q}^{10}-46\,{q}^{14} \Big) {A}^{26}+ \Big( 17+{q}^{-30}+{q}^{-26}+2\,{q}^{54}+5\,{q}^{48}+{q}^{38}-6\,{q}^{34}+10\,{q}^{42}+{q}^{60}-14\,{q}^{22}-3\,{q}^{20}-3\,{q}^{26}-{q}^{44}+10\,{q}^{-6}-{q}^{50}-5\,{q}^{40}-19\,{q}^{-2}-31\,{q}^{16}+{q}^{52}+11\,{q}^{36}-22\,{q}^{4}+25\,{q}^{24}+5\,{q}^{32}-2\,{q}^{-14}+3\,{q}^{-12}+6\,{q}^{-10}+25\,{q}^{12}+15\,{q}^{8}+{q}^{-24}+8\,{q}^{30}+8\,{q}^{-4}+2\,{q}^{-16}+11\,{q}^{2}+17\,{q}^{6}+30\,{q}^{18}+{q}^{-36}-10\,{q}^{-8}-5\,{q}^{28}-{q}^{-22}+2\,{q}^{-20}+{q}^{-18}-31\,{q}^{10}+15\,{q}^{14} \Big) {A}^{24}
$
}

\bigskip

Clearly, in this form they are not too informative.
A list, suitable for further studies is provided in the form of a txt-file at site
\cite{knotebook}.
Much more informative are expressions for the evolution families,
of which we provide a couple of simple
examples in the next section, and spectacularly elegant {\it properties}
of these complicated expressions in the later sections.

\section{Knot polynomials from evolution
\label{evomn}}

\subsection{$(m_1,n_1|m_2,n_2)$-evolution}

Studying the knot polynomials for {\it families} of knots is the most natural and effective approach,
see  \cite{DMMSS,evo,AENV,MMfing,MMMS21,MMMRSS} for motivation and examples.
The simplest  of this kind is the {\it evolution} method of \cite{DMMSS} and \cite{evo}, for the
three-strand braids $(m_1,n_1|m_2,n_2|\ldots)$ it studies the dependence on parameters $m_i$ and $n_i$.

The simplest family $(m,n)$ fully consists of {\it composites} of the 2-strand knots,
this means that their reduced polynomials are products of those for the
constituents, which in this case are the 2-strand torus knots/links with the HOMFLY
in arbitrary representation given in sec.\ref{2strand}:
\be
H_R^{(m,n)} = H_R^{(m)}\cdot H_R^{(n)}
\ee
Since knots/links $(m_1,n_1|m_2,0)=(m_1+m_2,n_1)$, the same remains true
for three-parametric families.

The first non-trivial family is $(m_1,n_1|m_2,n_2)$.
According to the general rules of the evolution method
\be\label{evoH}
H_R^{(m_1,n_1|m_2,n_2)} =
\sum_{a_1,a_2,a_3,a_4 = 1}^{N_R} \mathfrak{h}^{a_1a_2a_3a_4}_R
\lambda_{a_1}^{m_1} \lambda_{a_2}^{n_1} \lambda_{a_3}^{m_2} \lambda_{a_4}^{n_2}
\ee
where $N_R$ is the number of different representations in the square $R\otimes R$
(if the same representation appears in the symmetric and antisymmetric squares, it
contributes twice) and $\lambda_a$ are the corresponding ${\cal R}$-matrix eigenvalues,
$\lambda_Q = \epsilon_Q q^{\varkappa_Q-4\kappa_R}A^{-|R|}$ (the dependence on $R$ is fixed by the topological framing).
Thus, within the evolution method, the colored HOMFLY polynomials for our family  are described
by the 4-rank tensors $h_R$.

As a simple illustration,
in the fundamental representation $N_{[1]}=2$ $\ (Q=[2]_+\,,[1,1]_-)$ and
\be
\mathfrak{h}_{[1]}^{abcd} = \frac{\{Aq\}\{A/q\}}{[2]^4\{q\}^2}\cdot
\left\{\begin{array}{cccl}
\frac{\{Aq^3\}+2\{Aq\}}{\{A/q\}} & {\rm for} & abcd=&1111 \\
1 & {\rm for} & abcd = &1112,1121,1211,2111 \ {\rm and}\   1222,2122,2212,2221\\
-1 & {\rm for} & abcd = & 1122,1221,2211,2112 \\
\l[3] & {\rm for} & abcd = & 1212, 2121 \\
\frac{\{A/q^3\}+2\{A/q\}}{\{Aq\}} & {\rm for} & abcd=&2222
\end{array}
\right.
\ee
look rather elegant.

A lot of this structure survives in
the first non-symmetric representation, where  $N_{[2]}=3$ $\ (Q=[4]_+\,,[3,1]_-\,,[2,2]_+)$ and

\bigskip

{\footnotesize
\centerline{
$
\mathfrak{h}_{[2]}^{abcd} =
\frac{\{Aq^3\}\{Aq^2\}\{A/q\}}{\{q\}^4[4]^3[3]^3[2]^2} \cdot
\left\{\begin{array}{cccl}
\frac{[2]^4}{[4][3]\{A/q\}}
\cdot \left(\{Aq^6\}\{Aq^7\} +\{Aq^3\}\Big(2\{Aq^8\}+5[3]\{Aq^4\}+\{A\}\Big)\right)
& {\rm for} & abcd=&1111 \\
2{[3]^4[4]^2\over \{Aq^3\}}\{Aq^2\}^2+{[2]^4[3]^4\over [4]}\left(2{[4]\over [2]}(\{A\}-2\{Aq^2\})+[2]\{Aq\}+2\{Aq^6\}-\{A/q^4\}\right)   & {\rm for} & abcd=&2222 \\
\frac{\{A\}}{[4]^3[3][2]\{Aq^3\}\{Aq^2\}}
\cdot \left(\{A\}^2+[2]\{Aq^2\}\Big(2\{Aq\}+2\{A/q\}+\{A/q^3\}\Big)\right)
& {\rm for} & abcd=&3333 \\
\\
\l[2]^4\cdot \Big(2[2]\{Aq^3\}+\{Aq^6\}\Big) & {\rm for} & abcd = &1112,1121,1211,2111 \\
\l[2]^4\cdot [3]\Big(-[2]\{ Aq^3\}+\{A\} \Big) & {\rm for} & abcd = & 1122,1221,2211,2112 \\
\l[2]^4\cdot [3]\Big([2]\{Aq^7\}+\{Aq^4\}+[2]^2\{A\}\Big) & {\rm for} & abcd = & 1212, 2121 \\
\l[2]^4\cdot [3]^2\Big(2\{Aq^2\}-2\{A\}+\{A/q^2\}\Big)& {\rm for} & abcd = & 1222,2122,2212,2221\\
\\
 \frac{\{A\}  }{[3] }\cdot
\l[2]^3 & {\rm for} & abcd = &1113,1131,1311,3111    \\
 \frac{\{A\}  }{[3] }\cdot
\l[4][2]^2 & {\rm for} & abcd = & 1133,1331,3311,3113 \\
\frac{\{A\} }{[3] }\cdot
\l[5][4]^2[2]  & {\rm for} & abcd = & 1313, 3131 \\
 \frac{\{A\} }{[3] }\cdot
\l[4]^2[2] & {\rm for} & abcd = &  1333,3133,3313,3331\\
\\
\frac{[3]^2[2]^2 \{A\}}{\{Aq^3\}}  \Big(\{Aq^4\}-2\{Aq^2\}+3\{A\}-\{A/q^2\}+\{A/q^4\}\Big)
& {\rm for} & abcd = &2223,2232,2322,3222\\
\frac{[4][3][2]\{A\}}{\{Aq^3\}}\Big(\{Aq^2\}-2\{A\}-\{A/q^4\}\Big)
& {\rm for} & abcd = & 2233,2332,3322,3223 \\
\frac{[4]^2[3]\{A\}}{\{Aq^3\}}\Big(2\{A\}+\{A/q^2\}+\{A/q^4\}\Big)
 & {\rm for} & abcd = & 2323, 3232 \\
\frac{[4]^2\{A\}}{\{Aq^3\}}\Big(2\{A\}+\{A/q^2\}+\{A/q^4\}\Big)  & {\rm for} & abcd = &    2333,3233,3323,3332 \\
\\
-\l[2]^3\{A\} & {\rm for} & abcd = &1123,1231,2311,3112 \\
&&& {\rm and}\   1132,1321,3211,2113\\
\l[2]^3[5]\{A\} & {\rm for} & abcd = & 1213,2131,1312,3121 \\ \\
-[2]^2[6]\{A\}& {\rm for} & abcd = &1223,2231,2312,3122 \\
&&& {\rm and}\   3221,2213,2132,1322\\
\l[2]^3[3] \{A\}& {\rm for} & abcd = & 1232,2321,3212,2123 \\ \\
-\l[2]^2[4]\{A\}& {\rm for} & abcd = &1233,2331,3312,3123 \\
&&& {\rm and}\   2133,1332,3321,3213\\
\l[2]\{A\}& {\rm for} & abcd = & 1323,3231,2313,3132
\end{array}
\right.
$
}
}

\subsection{The family $(m,-1\,|\,1,-1)$
\label{evoser}}

\subsubsection{Generalities}

What we can provide at the current stage is
the answer for the simplest {\it one}-parametric family
$(m_1,n_1|m_2,n_2)=(n,-1\,|\,1,-1)$.
For the low odd values of $n$ it includes:
\be
\begin{array}{ccc|ccc}
(-1,-1\,|\,1,-1) & {\rm unknot} &&& (1,-1\,|\,1,-1) & 4_1   \\
(-3,-1\,|\,1,-1) & 5_2&&& (3,-1\,|\,1,-1) & 6_2    \\
(-5,-1\,|\,1,-1) & 7_3&&&(5,-1\,|\,1,-1) & 8_2    \\
 (-7,-1\,|\,1,-1) & 9_3&&&(7,-1\,|\,1,-1) & 10_2   \\
\ldots
\end{array}
\ee
The family $(n,-1\,|\,1,-1)$ describes the simple subfamily of pretzel knots $(n,\bar 2,1)$ at odd $n$ (see \cite{MMSpret}), while for even $n$ we get quite interesting two-component links.
This is next to the twist knots series in the Rolfsen table.
The twist knots ($\ldots,9_2,7_2,5_2,3_1,4_1,6_1,8_1,10_1,\ldots$)
are currently the main source of intuition about colored the
knot polynomials \cite{IMMMfe,sympols,evo},
and consideration of the next family is both natural and important.

Since the family under consideration belongs to the pretzel knots, the coefficients $h_{R,Y}(A,q)$ in the general evolution formula,
\be
H_R^{(n,-1\,|\,1,-1)}(A,q) = \sum_{Y\in R^{\otimes 2}}
\left(\frac{\epsilon_Y q^{\varkappa_Y}}{q^{4\varkappa_R}A^{|R|}}\right)^n
h_{R,Y}(A,q)
\label{serexpan}
\ee
are directly described in terms of the Racah matrices \cite[eq.(45)]{MMSpret}, \cite[eqs.(38)-(39)]{MMMRS}, which are known for all (anti)symmetric representations \cite{Racah,MMSpret}, as well as for representation $[2,1]$ (see \cite{GJ,MMMRS}):
\be\label{evop}
H_R^{(n,-1\,|\,1,-1)}(A,q)=H_R^{Pr(n,1,\bar 2)}(A,q)=d_R^2\sum_{X\in R\otimes\bar R}{1\over\sqrt{d_X}} \Big(ST^nS^\dagger\Big)_{\emptyset X} \Big(STS^\dagger\Big)_{\emptyset X}\Big(\bar S^\dagger\bar T^2\bar S\Big)_{\emptyset X}=\nn\\ =
d_R^2\sum_{Y\in R^{\otimes 2}}
\left(\frac{\epsilon_Y q^{\varkappa_Y}}{q^{4\varkappa_R}A^{|R|}}\right)^n\cdot\underbrace{\left(S_{\emptyset Y}\sum_{X\in R\otimes\bar R}{S_{YX}^\dagger\over\sqrt{d_X}} \Big(STS^\dagger\Big)_{\emptyset X}\Big(\bar S^\dagger\bar T^2\bar S\Big)_{\emptyset X}\right)}_{h_{R,Y}(A,q)}
\ee
where $S$ is the Racah matrix that relates the cases of differently placed brackets in the map of the product: $R\otimes R\otimes\bar R\to R$, while $\bar S$ for $R\otimes \bar R\otimes R\to R$, and $T$ is the diagonal matrix with entries being eigenvalues $\frac{\epsilon_Y q^{\varkappa_Y}}{q^{4\varkappa_R}A^{|R|}}$. Note that $S_{\emptyset X}=\sqrt{d_X}/d_R$ and $\bar S_{\emptyset X}=\sqrt{d_X}/d_R$.

Thus, we list below the coefficients $h_{R,Y}(A,q)$ for the (anti)symmetric representations and representation $[2,1]$ and calculate for the $[3,1]$ case from the 3-strand representation of this paper, since the proper Racah matrices are not known yet.

After summation over $X$ or from the coefficients $\mathfrak{h}^{a_1a_2a_3a_4}_R$ of (\ref{evoH}) after
weighted summation (averaging) over three indices $a_{2,3,4}$, formula (\ref{evop}) can be rewritten in the form of corrections to
the 2-strand formula (\ref{2straH}),
in terms of the $A$-independent coefficients $C^{R,Y}_i\!\!(q)$:
\be
H_R^{(n,-1\,|\,1,-1)}(A,q) = (Aq^{\mu_R})^{2|R|}\cdot
 \underbrace{\sum_{Y\in R\otimes R} \frac{d_Y}{d_R}
 \cdot \left(\frac{\epsilon_Y q^{\varkappa_Y}}{q^{4\varkappa_R}A^{|R|}}\right)^n
 \cdot\ \ \ \ \   }_{H^{(n)}_R}
 \!\!\!\!\!\!\!\!\!\!\!
\left(1 + \sum_{i=1}^{|R|}  (-)^i\,\,C^{R,Y}_i\!\!(q)\cdot
\left(\frac{\{q\}}{A^2q^{2\mu_R}}\right)^i\right)
\label{m111H}
\ee
where $\mu_R = r-s$ for the single hook representations $R=[r,1^{s-1}]$.

\subsubsection{Symmetric representations $R=[r]$}

For symmetric $R=[r]$ the sum in (\ref{serexpan}) goes over the representations $[2r-a,a]$, and one can use formulas from \cite{MMSpret,Racah} to write manifestly
\be
h_{[r],[2r-a,a]}(A,q)=(-1)^{a+1}\sum_{k=0}^r \bar d_k \mathfrak{a}_{ka}\cdot\Big(\sum_{i=0}^r\mathfrak{a}_{ki}q^{i^2+i}\Big) \Big(\sum_{j=0}^r\bar{\mathfrak{a}}_{kj}q^{2j^2-2j}A^{2j}\Big)
\label{symrepsser}
\ee
where $\bar d_k$ is the quantum dimension of the representations arising in the decomposition $[r]\otimes\overline{[r]} = \oplus_{k=0}^r \ [2k,k^{N-2}]$:
\be
\bar d_k=D_{2k-1}\cdot \left(\prod_{j=0}^{k-2}\frac{D_j}{[j+2]} \right)^{\!\!2}\!\cdot D_{-1}
\ee
and
\be
\mathfrak{a}_{km} = (-1)^{r+k} [2m+1]{G(r-m)\over G(r+k+1)}\cdot {\Big([k]![m]!\Big)^2\,[r-k]!\,[r-m] !\over [r+k+1]!\,[r+m+1]!}
 \times \nn \\ \times
\ \sum_{j=\text{max}(r+m,r+k)}^{\text{min}(r+k+m,2r)} {(-1)^j\,[j+1]!\over [2r-j]!\ \Big([j-r-k]!\,[j-r-m]!\,[r+k+m-j]!\Big)^2} \cdot \frac{G(j+1)}{G(j-r-m)}\nn\\
\bar{\mathfrak{a}}_{km} = {(-1)^{r+k+m} D_{2m-1}G(m)^2\over G(r+k+1)\,G(r+m+1)}\cdot {\Big([k]![m]!\Big)^2\,[r-k]!\,[r-m] !\over [r+k+1]!\,[r+m+1]!}
 \times \nn \\ \times
\ \sum_{j=\text{max}(r+m,r+k)}^{\text{min}(r+k+m,2r)} {(-1)^j\,[j+1]!\over [2r-j]!\ \Big([j-r-k]!\,[j-r-m]!\,[r+k+m-j]!\Big)^2} \cdot \frac{G(j+1)}{G(r+k+m-j)}
\ee
with
\be
G(n) &=& \dfrac{1}{[n]!}\prod_{i=-1}^{n-2}D_{i}={(A/q;q)_n\over (q;q)_n}
\ee
where we used the symmetric q-Pochhammer symbol $(A;q)_n=\prod_{j=0}^{n-1}\{Aq^j\}$. At $A=q^N$, $G(n)$ becomes the q-binomial
$\left(\begin{array}{c} N+n-2\\n
\end{array}\right)_q$.

From (\ref{symrepsser}) one can also read off the coefficients $C_i(r,a)$ in representation of $h(A,q)$ in form (\ref{m111H})
\be
h_{[r],[2r-a,a]}(A,q)
=\big(q^{r-1}A\big)^{2r}\cdot
\frac{d_{[2r-a,a]}}{d_{[r]}}
\cdot
\left(1 + \sum_{i=1}^r (-)^iq^{\frac{i(i+1)}{2}}[i]! \ C_i(r,a)
\left(\frac{\{q\}}{(q^{r-1}A)^2}\right)^i\right)
\ee
In particular, one can realize the following properties of these coefficients:

\begin{itemize}
\item They have a symmetry $i\longrightarrow r-i$,
\be
C_{r-i}(r,a|q) = (-)^a\cdot q^{-2ar+a(a-1)}\cdot C_i(r,a|q^{-1})
\label{symm}
\ee
involving also the change $q\longrightarrow q^{-1}$ and rescaling.
\item They satisfy simple sum rules in lines:
\be
\begin{array}{lcl}
\sum_{i=0}^r   C_i(r,r) = \prod_{j=1}^r (1-q^{-2j})
&& \sum_{i=0}^r (-)^i C_i(r,r) =\prod_{j=1}^r (1+q^{-2j})
=q^{-\frac{r(r-1)}{2}}\prod_{j=1}^r \frac{[2j]}{[j]}  \\
\sum_{i=0}^r   C_i(r,r-1) = 2\prod_{j=2}^r(1-q^{-2j})
&& \sum_{i=0}^r (-)^i C_i(r,r-1) = 0   \\
\sum_{i=0}^r   C_i(r,r-2) = (2+[2]^2)\prod_{j=3}^r(1-q^{-2j})
&& \sum_{i=0}^r (-)^i C_i(r,r-2) = -q^{-\frac{(r-2)(r+3)}{2}} \prod_{j=2}^r \frac{[2j]}{[j]}   \\
\sum_{i=0}^r   C_i(r,r-3) = 2(1+[3]^2)\prod_{j=4}^r(1-q^{-2j})
&& \sum_{i=0}^r (-)^i C_i(r,r-3) = 0   \\
\ldots    && \ldots
\end{array}
\ee
or, in general,
\be
\sum_{i=0}^r   C_i(r,r-b) =
q^{-\frac{(r-b)(r+b+1)}{2}}\,\{q\}^{r-b}\,\frac{[r]!}{[b]!}\cdot
\sum_{i=0}^b \left(\frac{[b]!}{[j]![b-j]!}\right)^2
\nn \\
\sum_{i=0}^r   (-)^i C_i(r,r-b) =
(-)^{\frac{b}{2}} q^{-\frac{(r-b)(r+b+1)}{2}}\, \frac{[2r]!! \, [b]!}{[r]!\, ([b]!!)^2}
\cdot\left\{
\begin{array}{ccc}  0 & {\rm for\ odd} & b \\
1 & {\rm for \ even} & b
\end{array}\right.
\ee
\end{itemize}

\subsubsection{Totally antisymmetric representations $R = [1^r]$}

As usual (rank-level duality \cite{DMMSS,GS,IMMMfe}), for antisymmetric representations
\be
H_{[1^r]}^{(n,-1\,|\,1,-1)}(A,q) =  H_{[r]}^{(n,-1\,|\,1,-1)}(A,q^{-1})
\ee

\subsubsection{Representation $R=[2,1]$
\label{21ser}}

In this case, the evolution coefficients can be calculated both from formula (\ref{evop}) and by the evolution method from several known knot polynomials \cite{evo}. Let us use the second possibility here, since only this approach can be used in the $[3,1]$ case.

For $[2,1]^{\otimes 2}$ the eigenvalues of ${\cal R}$-matrix are
{\footnotesize
\be
\lambda_{[4,2]}=\frac{q^5}{A^3}, \ \ \ \   \lambda_{[4,1,1]}=-\frac{q^3}{A^3}, \ \ \ \
\lambda_{[3,3]}=-\frac{q^3}{A^3}, \ \ \ \
\lambda_{[3,2,1]_\pm}=\pm \frac{1}{A^3}, \ \ \ \
\lambda_{[3,1,1,1]}=\frac{1}{q^3A^3}, \ \ \ \   \lambda_{[2,2,2]}=\frac{1}{q^3A^3}, \ \ \ \
\lambda_{[2,2,1,1]}=-\frac{1}{q^5A^3}
\nn
\ee
}
and therefore the $m$-evolution is defined as follows:
\be
\!\!\!H_{[2,1]}^{(n,-1\,|\,1,-1)}\! =
\frac{h_{[4,2]} q^{5n}+\Big(h_{[4,1,1]}+h_{[3,3]}\Big)(-q^3)^n+h_{[3,2,1]_+}+h_{[3,2,1]_-}(-)^n
+\Big(h_{[3,1,1,1]}+h_{[2,2,2]}\Big) (q^{-3})^n +h_{[2,2,1,1]} (-q^{-5})^n }{A^{3n}}
\ee
Note that two pairs of representations have the same eigenvalues, thus their contributions
are not separated.
The two representations $[3,2,1]_\pm$ contribute equivalently to knot polynomials:
for odd $n$ only the difference $h_{[3,2,1]_+}-h_{[3,2,1]_-}$ is seen,
but link polynomials with even $n$ separate these two contributions.
The other two sums are inseparable within the $(n,-1\,|\,1,-1)$ series,
we, however, suggest a plausible decomposition, which can be checked in the analysis
of richer evolution patterns. In fact, {\it after} proposing these decompositions we derived them using (\ref{evop}) and manifest formulas for the Racah matrices from \cite{MMMRS} (notice an additional sign factors that have to be added in this case in accordance with \cite[s.7]{MMMRS} and \cite[s.2.7]{MMMRSS}). Unfortunately, at the moment, these Racah matrices are not known for $[3,1]$ case yet and, hence, such a method is unavailable in that case. Finally, the values of the coefficients are:
\be
\begin{array}{rl}
\nn\\
\!\!\!\!\!\!\!\!\!\!
h_{[4,2]}=
 \ \ \ \ \ \ \ \ \ \ \ &\!\!\!\!\!\!\!\!\!\!\!\!\!\!\!\!\!\!\!\!
\frac{d_{[4,2]}}{d_{[2,1]}} \cdot
{\footnotesize
\Big(A^6 - q\cdot(q^4+q^2+2+q^{-4})\{q\}A^4 + q^4(q^4+q^2+3+q^{-2}+q^{-4})\{q\}^2A^2  - q^5[3]\{q\}^3\Big)
}
\nn \\ \nn \\
h_{[4,1,1]}+h_{[3,3]} \ =&
\frac{d_{[4,1,1]}}{d_{[2,1]}}  \cdot
{\footnotesize
\Big(A^6 - q\cdot (q^4+1+q^{-4})\{q\}A^4  -q^2(q^4+1+q^{-4})\{q\}^2A^2 + q^3[3]\{q\}^3\Big)
}  + \nn\\
&
 \!\!\!\!\!\!\!\!\!\!\!
+ \frac{d_{[3,3]}}{d_{[2,1]}}   \cdot
{\footnotesize
\Big(A^6 -   q\cdot(q^4+q^2+2-q^{-6})\{q\}A^4  + q^2(q^6+q^4+2q^2-q^{-4})\{q\}^2A^2  + q^3[3]\{q\}^3\Big)
}
\nn \\ \nn \\
\frac{1}{2}\Big(h_{[3,2,1]_+} + h_{[3,2,1]_-}\Big) =
&
\frac{d_{[3,2,1]}}{d_{[2,1]}}
{\footnotesize
\Big( A^6 - \frac{[3][4]}{[2]}\{q\}^2 A^4 -[3]\{q\}^2A^2\Big)
}
\nn \\ \nn \\
\frac{1}{2}\Big(h_{[3,2,1]_+} - h_{[3,2,1]_-}\Big)
=
&
\ \ \ \ \ \ \ \ \ \ \ \ \ \ -\frac{d_{[3,2,1]}}{d_{[2,1]}}\cdot
{\footnotesize
[3]\{q\}^3\Big(A^2(q^2+q^{-2})+1\Big)
}
\nn \\ \nn \\
h_{[3,1,1,1]}+h_{[2,2,2]} \ =&
\frac{d_{[3,1,1,1]}}{d_{[2,1]}} \cdot
{\footnotesize
\Big(A^6 +q^{-1}  (q^4+1+q^{-4})\{q\}A^4  -q^{-2}(q^4+1+q^{-4})\{q\}^2A^2 - q^{-3}[3]\{q\}^3\Big)
}  + \nn\\
&
\!\!\!\!\!\!\!\!\!\!\!  \!\!\!\!\!\!\!\!\!\!\!
+ \frac{d_{[2,2,2]}}{d_{[2,1]}} \cdot
{\footnotesize
\Big(A^6 +   q^{-1}(q^{-4}+q^{-2}+2-q^{6})\{q\}A^4  + q^{-2}(q^{-6}+q^{-4}+2q^{-2}-q^{4})\{q\}^2A^2
- q^{-3}[3]\{q\}^3\Big)
}
\nn \\ \nn \\
h_{[2,2,1,1]}  =
\ \ \ \ \ \ \ \ \ \ \  &\!\!\!\!\!\!\!\!\!\!\!\!\!\!\!\!\!\!\!\!\!\!\!\!
\frac{d_{[2,2,1,1]}}{d_{[2,1]}}    \cdot
{\footnotesize
\Big(A^6 + q^{-1}(q^4+2  +q^{-2}+q^{-4})\{q\}A^4 +q^{-4}(q^4+q^2+3+q^{-2}+q^{-4}) \{q\}^2A^2  +
q^{-5}[3]\{q\}^3\Big)
}
\nn\\
\end{array}
\nn
\ee
The diagram $[2,1]$ is symmetric, this explains the simple form of the coefficients
and their symmetries.
Clearly, for the transposed diagrams one has
$h_{[2,2,1,1]} (q)= h_{[4,2]}(-q^{-1})$,
$h_{[3,1,1,1]}(q) = h_{[4,1,1]}(-q^{-1})$,
$h_{[2,2,2]}(q) = h_{[3,3]}(-q^{-1})$
and $h_{[3,2,1]_\pm}(q) = h_{[3,2,1]_{\pm}}(-q^{-1})$.
Note that the sum and difference of $h_{[321]_\pm}$ have different parities of powers in $q$:
this is because the corresponding eigenvalue differs in this power from all the others and there
is a need to compensate for the difference between odd and even $m$ for knots and links.

\subsubsection{Representation $R=[3,1]$
\label{31ser}}

For $[3,1]^{\otimes 2}$ the ${\cal R}$-matrix eigenvalues
were already listed in (\ref{rev31}). In the topological framing, they are:
{\footnotesize
\be
\lambda_{[6,2]}=\frac{q^{14}}{q^8A^4}, \ \ \ \  \lambda_{[6,1,1]}=-\frac{q^{12}}{q^8A^4}, \ \ \ \
\lambda_{[5,3]}=-\frac{q^{10}}{q^8A^4}, \ \ \ \  \lambda_{[4,4]}=\frac{q^8}{q^8A^4}, \ \ \ \
\lambda_{[5,2,1]_\pm}=\pm\frac{q^7}{q^8A^4}, \ \ \ \ \lambda_{[5,1,1,1]}=\frac{q^4}{q^8A^4},
\nn \\
\lambda_{[4,3,1]_\pm}=\pm \frac{q^4}{q^8A^4}, \ \ \ \
\lambda_{[4,2,2]}=\frac{q^2}{q^8A^4}, \ \ \ \  \lambda_{[4,2,1,1]}=-\frac{1}{q^8A^4}, \ \ \ \
\lambda_{[3,3,2]}=-\frac{1}{q^8A^4}, \ \ \ \  \lambda_{[3,3,1,1]}=\frac{q^{-2}}{  q^8A^4}, \ \ \ \
\nn
\ee
}
and
\be
H_{[3,1]}^{(n,-1\,|\,1,-1)} = (A^4q^8)^{-n}\cdot \Big\{
h_{[6,2]} q^{14n}+h_{[6,1,1]}(-q^{12})^n+h_{[5,3]}(-q^{10})^n+h_{[44]} q^{8n}
+h_{[5,2,1]_+} q^{7n}+h_{[5,2,1]_-}(-q^{7})^{n} + \nn \\
+\Big(h_{[5,1,1,1]}+h_{[4,3,1]_+}\Big) q^{4n}
+h_{[4,3,1]_-}(-q^{4})^{n} +h_{[4,2,2]} q^{2n}
+\Big(h_{[4,2,1,1]}+h_{[3,3,2]}\Big)\cdot (-1)^n+h_{[3,3,1,1]} q^{-2n}\Big\}
\ee
Our evaluation of $[3,1]$-colored HOMFLY, together with additional arguments
already used in sec.\ref{21ser}, allows one to find/conjecture all these coefficients
$h^{(\cdot,-1\,|\,1,-1)}$:

{\footnotesize
\be
\begin{array}{rll}
h_{[6,2]} = &
\frac{d_{[6,2]}}{d_{[3,1]}}\,\Big\{ (qA)^8
-  \{q\}(qA)^6 (q^3+q^{-3})(q^4+q^2+2+q^{-2})\
&  + [2]\{q\}^2(qA)^4(q^{11}+3q^7+2q^5+2q^3+3q+2q^{-1}+q^{-5}) -   \nn \\
&&
-  [2]^2\{q\}^3(qA)^2(q^{11}+q^9+q^7 +3q^{5} +q)\ +\ q^8[4][2]\{q\}^4    \Big\}
\nn \\ \nn \\
h_{[6,1,1]} =&
\frac{d_{[6,1,1]}}{d_{[3,1]}}\,\Big\{ (qA)^8
- [4]\{q\}(qA)^6 (q^4-q^2+1+q^{-2})

& - [2]\{q\}^2(qA)^4(q^9-q^7-q^3-q-q^{-1}-q^{-5})
+   \nn \\
&&+   [2]^2\{q\}^3(qA)^2(q^9+q^7+2q^3-q+q^{-1})\ -\ q^6[4][2]\{q\}^4    \Big\}
\nn \\ \nn \\
h_{[5,3]} = &
\frac{d_{[5,3]}}{d_{[3,1]}}\,\Big\{ (qA)^8
 - [2]\{q\}(qA)^6 (q^6+2q^2+q^{-4}- q^{-6})

&+ [2]\{q\}^2(qA)^4(q^{11}+2q^7+2q^5+q+q^{-1}-2q^{-3}-q^{-9})
-   \nn \\
&&-   [2]^2\{q\}^3(qA)^2(q^9-q^7+q^5-2q+q^{-1}-q^{-3})\ -\ q^4[4][2]\{q\}^4    \Big\}
\nn \\ \nn \\
h_{[4,4]} = &
\frac{d_{[4,4]}}{d_{[3,1]}}\,
\Big\{ (qA)^8
- [4]\{q\}(qA)^6 (q^{4}+1-q^{-4})
&+ [4]\{q\}^2(qA)^4(q^{9}+q^{5}+q^{3}-q-q^{-1}-q^{-5}+q^{-9})
+ \nn \\
&&+ [2][4]\{q\}^3(qA)^2(q^{5}+q^{-1}-q^{-3})
+ [2][4]\{q\}^4 q^2    \Big\}
\nn \\ \nn \\
\!\!\!\!\!\!\!\!\!\!\!\!\!\!\!
\frac{1}{2}\Big(h_{[5,2,1]_+} + h_{[5,2,1]_-}\Big) = &
\frac{d_{[5,2,1]}} {d_{[3,1]}} \, \Big\{(qA)^8 -
\Big(q^7+q^3+q+q^{-3}-q^{-7}\Big)\{q\}(qA)^6
&+\Big(q^6-2q^{-2}-2q^{-4}-q^{-6}-2q^{-8}-q^{-10}\Big)\{q\}^2(qA)^4+\nn \\
&& +q^{-1} [2][4] \{q\}^3(qA)^2
\Big\}
\nn \\ \nn \\
\!\!\!\!\!\!\!\!\!\!\!\!\!\!\!\!\!\!\!\!\!\!\!\!\!\!\!\!\!\!
\frac{1}{2}\Big(h_{[5,2,1]_+} - h_{[5,2,1]_-}\Big) =
&\ \ \ \ \ \ \ \ \ \ \ \ \ \ \ \ \ \ \ \ \ \ \ \ \ \ \ \ \ \ -\frac{d_{[5,2,1]}} {d_{[3,1]}} \cdot
[2][4]\{q\}^3\cdot
& \!\!\!\!\!\!\!\!\!\!\!\!\!\!\!\!\!\!\!\!\!\!\!\!\!\!\!\!\!\!\!\!\!\!\!\!\!\!\! \Big\{
q(q^3+q^{-3}) (qA)^4+ (qA)^2\Big(-q^4+q^2+1-q^{-2}+q^{-4}\Big) -q\{q\}\Big\}
\nn \\ \nn \\
\!\!\!\!\!\!\!\!\!\!\!\!\!\!\!\!\!\!\!\!\!\!\!\!\!\!\!\!\!\!
h_{[5,1,1,1]}+h_{[4,3,1]_+} \ \stackrel{?}{=}&
\frac{d_{[5,1,1,1]}}{d_{[3,1]}}\,\Big\{(qA)^8 + q^{-1}[2][4]\{q\}^3(qA)^6
& \!\!\!\!\!\!\!\!\!\!\!\!\!\!\!\!\!\!\!\!\!\!\!\!\!\!\!\!\!\!
- [2]^2[4]\{q\}^2(qA)^4(q-q^{-1}+q^{-5})
+ q^{-2}[2]^2[4]\{q\}^5(qA)^2+q^{-2}[2][4]\{q\}^4 \Big\} + \nn \\ \nn \\
&\!\!\!\!\!\!\!\!\!\!\!\!\!\!\!\!\!\!\!\!\!\!\!\!\!\!\!\!\!\!\!\!\!\!\!\!\!\!\!\!\!\!
+ \frac{d_{[4,3,1]}}{d_{[3,1]}} \,\Big\{(qA)^8 -[4]\{q\}^2(qA)^6(q^3+q+q^{-3})
&\!\!\!\!\!\!\!\!\!\!\!\!\!\!\!\!\!\!\!\!\!\!\!\!\!\!\!\!\!\!\!\!\!\!\!\!\!\!\!\!\!\!\!\!\!\!\!\!\!\!\!
\!\!\!\!\!\!\!\!\!\!\!\!\!\!\!\!\!\!\!\!\!
+[4]\{q\}^2(qA)^4(q^5-q^3-1-q^{-5}+q^{-9}) +[2][4]\{q\}^4(qA)^2(1+q^{-2}+q^{-6})+q^{-2}[4][2]\{q\}^4 \Big\}
\nn \\ \nn \\
h_{[4,3,1]_-} =&
\frac{d_{[4,3,1]}}{d_{[3,1]}}\,\Big\{(qA)^{8}
- [4]\{q\}^2(qA)^6(q^{3}+q^{-1}+q^{-3})
&\!\!\!\!\!\!\!\!\!\! - [4]\{q\}^2(qA)^4(q^{5}-q^{3}+q+2q^{-1}-2q^{-3}+q^{-5}-q^{-9})
-\nn \\
&&- [4]\{q\}^3(qA)^2(q^{2}+1-q^{-2}+2q^{-4}-q^{-8})
- [2][4]\{q\}^4 q^{-2}
\Big\}
\nn\\ \nn \\
h_{[4,2,2]} =&
\frac{d_{[4,2,2]}}{d_{[3,1]}}\,\Big\{ (qA)^8
- (q^3+q^{-3})\{q\}(qA)^6 (q^4-q^{-2}- q^{-4})
&+ [2]\{q\}^2(qA)^4(q^{5}-q^3-2q^{-3}+q^{-5}-q^{-7}+q^{-11})
+   \nn \\
&&
\!\!\!\!\!\!\!\!\!\!\!\!\!\!\!\!\!\!\!\!\!\!\!\!\!\!\!\!\!\!\!\!\!\!\!\!\!\!\!\!\!\!\!\!\!\!\!
+  [2]\{q\}^3(qA)^2( q^2+1-q^{-2}-3q^{-6}-q^{-8}-q^{-10}-q^{-12})\ +\ q^{-4}[4][2]\{q\}^4 \Big\}
\nn \\ \nn \\
\!\!\!\!\!\!\!\!\!\!\!\!\!\!\!\!\!\!\!\!\!\!\!
h_{[4,2,1,1]}+h_{[3,3,2]} \ \stackrel{?}{=} &
\frac{d_{[4,2,1,1]}}{d_{[3,1]}}\,\Big\{ (qA)^8 + \frac{q^6-q^4+1}{q^4}\,[4]\{q\}(qA)^6
- \frac{(q^6-q^4+1)\{q\}}{q^7}\,[4][2]\{q\}^2(qA)^4
\!\!\!\!\!\!\!\!\!\!\!\!\!\! \!\!\!\!\!\!\!\!\!\!\!\!\!\! \!\!\!\!\!\!\!\!\!\!\!\!\!\!
&\ \ \ \ \ \ \ \ \ \ \ \ \ \ \ \ \ \
-\frac{q^8+q^6-q^4+q^2+1}{q^{10}}\,[4]\{q\}^3(qA)^2 -q^{-6}[4][2]\{q\}^4 \Big\} + \nn \\ \nn \\
& + \frac{d_{[3,3,2]}}{d_{[3,1]}}\,\Big\{ (qA)^8 - \frac{q^8-q^6+1}{q^4}\,[4]\{q\}(qA)^6
- \frac{q^8-q^6-1 }{q^8}\,[4][2]\{q\}^2(qA)^4
\!\!\!\!\!\!\!\!\!\!\!\!\!\! \!\!\!\!\!\!\!\!\!\!\!\!\!\! \!\!\!\!\!\!\!\!\!\!\!\!\!\!
&\ \ \ \ \ \ \ \ \ \ \ \ \
 -\frac{q^{10}+q^8-q^6 -q^2-1}{q^{12}}\,[4]\{q\}^3(qA)^2 -q^{-6}[4][2]\{q\}^4\Big\}
\nn \\ \nn \\
h_{[3,3,1,1]} = &
\frac{d_{[3,1,1,1]}}{d_{[3,1]}}\,\Big\{ (qA)^8
+[2]\{q\}(qA)^6 (q^4-q^2+1+q^{-2}+ q^{-6})
&+ [2]\{q\}^2(qA)^4(q +q^{-3}+2q^{-5}+q^{-7}+q^{-9}+q^{-11})
+   \nn \\
&& \!\!\!\!\!\!\!\!\!\!\!\!\!\!\!\!\!\!\!\!\!\!\!\!\!\!\!\!\!\!\!\!\!\!\!\!\!\!\!\!\!\!\!\!\!\!\!\!
+  [2]\{q\}^3(qA)^2\cdot  q^{-2}(1+q^{-2}+q^{-6})(1+q^{-4}+q^{-6}) \ +\ q^{-8}[4][2]\{q\}^4    \Big\}
\end{array}
\ee
}
where the questions marks means conjectural decompositions of the sums. As a corollary of these formulas

{\footnotesize
\be
\frac{1}{2}\Big(h_{[4,3,1]_+} + h_{[4,3,1]_-}\Big) \ \stackrel{?}{=}&
\frac{d_{[4,3,1]}}{d_{[3,1]}}\,\Big\{(qA)^{8}
- [4]\{q\}^2(qA)^6(q^{3}+q^{-1}+q^{-3})
  - [4]\{q\}^2(qA)^4( q+q^{-1}-q^{-3}+q^{-5}-q^{-9})-q^{-4}[4]\{q\}^3(qA)^2\Big\}
\nn \\ \nn \\
\frac{1}{2}\Big(h_{[4,3,1]_+} - h_{[4,3,1]_-}\Big) \ \stackrel{?}{=}&
\frac{d_{[4,3,1]}}{d_{[3,1]}}\cdot [4]\{q\}^3\cdot \Big\{ q(q^3+q^{-3})(qA)^{4}
+ (qA)^2(q^2+1-q^{-2}+q^{-4}-q^{-8}) + q^{-2}[2]\{q\}\Big\}
\nn\\
\ee
}

\section{Generic properties of $H_{[3,1]}$ \label{proppols}}

\subsection{Special and Alexander polynomials at $q=1$ or $A=1$}
\label{spal}

According to \cite{DMMSS,IMMMfe,Che,anton,GGS},
\be
H_R(A) = H_1^{|R|}(A) \ \ \ \ {\rm at} \ \ q=1 \ \ \forall \ {\rm representations}\ R
\label{spepo}
\ee
and
\be
H_R(q) = H_1\Big(q^{|R|}\Big) \ \ \ \ {\rm at} \ \ A=1 \ \ \forall \ {\rm single hook\ representations}\
R = [r,1^{s-1}]
\ee
This holds for all our 3-strand $H_{[3,1]}$.

\subsection{Factorization at roots of unity, $q^8=1$}

According to \cite{Konfact},
\be
H_{[3,1]}=H_{[4]} \ \ \ \ {\rm at} \ \ q^8=1
\ee
This is true in all our examples and this provides a new serious support to
the factorization conjectures of \cite{Konfact}.

\subsection{Symmetric Jones polynomials}

It is well-known that the representation $R=[3,1]$ reduces to $R=[2]$ for the Lie algebra $sl_2$. Therefore, if we put $A=q^2$, the HOMFLY polynomial $H_{[3,1]}$ reduces to the Jones polynomial in the first symmetric representation $J_{[2]}$:
\be
H_{[3,1]} = J_{[2]} \ \ \ \ {\rm at} \ \ A=q^2.
\ee
Since the symmetric Jones polynomials are known for many knots, in particular, for all knots from the Rolfsen table, one can use them to check our results.

\subsection{Universality and adjoint representation at $A=q^4$}

By the general rule (rank-level duality \cite{DMMSS,GS,IMMMfe}), for the transposition of Young diagram
\be
H_{R^{tr}}(A,q) = H_R(A, -{q}^{-1})
\label{trans}
\ee
(an additional inversion $A\to -A^{-1}$ provides a mirror knot
instead of transposing the representation).

Thus, together with $H_{[3,1]}$ we simultaneously know
\be
H_{[2,1,1]}(A,q) = H_{[3,1]}(A,-q^{-1})
\label{trans211}
\ee
For the particular case of $sl_4$ algebra, $[2,1,1]$ is the adjoint representation
and the adjoint HOMFLY polynomial satisfies \cite{MMkrM,univev} the {\it universality} hypothesis \cite{Vog},
unifying them with the adjoint polynomials for other groups,
including the much simpler adjoint Kauffman polynomials
(they are simpler because the adjoint representation of $so_N$ is just $[1,1]$
for all $N$, while it is an $N$-dependent $[2,q^{N-2}]$ for $sl_N$).
This allows one to compare our $H_{[3,1]}$ at $A=q^4$ with the universal
formulas from \cite{univev}. They match in all examples that we looked at.

\subsection{Loop expansion and Vassiliev invariants}

The HOMFLY polynomials have an interesting well-defined expansion when
$\hbar\to 1, \ A=\exp\left(\frac{N\hbar}{2}\right), \ q=\exp\left(\frac{\hbar}{2}\right)$.
This expansion is known as loop expansion,
and the coefficients are the celebrated Vassiliev invariants \cite{vassinv}.
In the Vassiliev approach, the HOMFLY polynomial can be written as \cite{DBSS}
\be\label{vas}
H_R^{\cal K}(A=e^{\frac{N\hbar}{2}}|q=e^{\frac{\hbar}{2}}) =
\sum_{i=0}^{\infty}\hbar^{i} \sum_{j=1}^{\mathcal{N}_i}  r_{i,j}^{(R)} v_{i,j}^{\cal K}
\ee
where $r_{i,j}^{(R)}$ are the polynomials of degree $|i|$ in $N$ corresponding
to the trivalent diagrams \cite{Labastida,DBSS},
and $\mathcal{N}_i$ is the dimension of the vector space formed by the trivalent diagrams.
Here $v_{i,j}^{\cal K}$ are finite type knot invariants introduced by V.Vassiliev.
Thus, what stands in (\ref{vas}) is the double series in powers of $\hbar$ and $N$,
such that the degree of $\hbar$ exceeds or equals to the degree of $N$. The polynomials $r_{i,j}^{(R)}$
are known up to degree $6$ in arbitrary representation $R$,
whereas the Vassiliev invariants up to order $6$ are known for all knots with number of crossings less than 15.
Thus, it provides a lot of explicit checks.
To be more concrete, we consider one example of the knot $10_{161}$ from s.\ref{exaknots}.

\bigskip

As an illustration, we look at the first terms of the expansion (\ref{vas})
in the case of $R=[3,1]$ and for the knot $10_{161}$ mentioned  in sec.\ref{exaknots}.
Taking the knot-independent trivalent diagrams  from \cite{MMSS}
and the lowest Vassiliev invariants for $10_{161}$ from \cite{katlas},
{\footnotesize
\be\label{10161}
\begin{array}{ccl}
r_{2,1} = -N^2-N+4 & \ \ \ \ \ \ \ & v_{2,1}^{10_{161}} = 28 \\ \\
r_{3,1} = {1\over2}\,N\,( \,N^2+N-4 ) & & v_{3,1}^{10_{161}} = -144 \\ \\
r_{4,1} = \left( -{N}^{2}-N+4 \right)^{2} & & v_{4,1}^{10_{161}} =  392 \\ \\
r_{4,2} = -{1\over4}\,{N}^{2} \left( \,{N}^{2}+\,N-4 \right) & & v_{4,2}^{10_{161}} = {2882 \over 3} \\ \\
r_{4,3} = {1\over4}\,{N}^{4}+{3\over4}\,{N}^{3}+{11\over2}\,{N}^{2}+4\,N-32 & &
v_{4,3}^{10_{161}} = {430 \over 3}
\end{array}
\ee
}

\noindent
we obtain the first terms of the $\hbar$-expansion (\ref{vas}):
{\footnotesize
\be
H_{[3,1]}^{10_{161}} = 1- 28\left( {N}^{2}+N-4 \right) {\hbar}^{2}
-72N \left( {N}^{2}+N-4 \right) {\hbar}^{3}+ \left( {\frac {5056}{3}}-995{N}^{2}-{\frac {7688N}{3}}
+ {\frac{1954{N}^{3}}{3}} + {\frac {563{N}^{4}}{3}} \right) {\hbar}^{4} + O \left( {\hbar}^{5} \right)
\nn
\ee
}
Comparing with the coefficients of the $\hbar$ expansion of (\ref{10161}), we find a complete agreement.

At the present moment, we know the trivalent diagrams explicitly only up to degree $6$
(they are available at \cite{MMSS} up to degree $4$).
The corresponding Vassiliev invariants are fully determined by the HOMFLY polynomials
in representations $R=[1]$ and $R=[2]$.
To extract new Vassiliev invariants from more interesting higher symmetric, $[2,1]$-
and the newly-calculated $[3,1]$-colored HOMFLY,  more complicated diagrams are needed.

\subsection{Genus expansion and Hurwitz $\tau$-function}

Another interesting expansion of the HOMFLY polynomials is the large $N$,
or {\it genus} expansion,
which relates knot theory with the Hurwitz enumeration problem.
In this expansion, one expands the knot polynomials in powers of $\hbar$ ($q=e^\hbar$) similarly to the previous subsection, but now $A$, i.e. $\hbar \cdot N $ rather that $N$ is fixed.
It turns out that the Ooguri-Vafa partition functions of the colored HOMFLY polynomials \cite{OV} is equal
to certain generating functions of the Hurwitz numbers \cite{HurwMMS,MMSS}, hence,
we also call this expansion Hurwitz, and the corresponding generalization of the
KP/Toda $\tau$-functions (a general solution to the AMM/EO topological recursion \cite{AMM/EO})
is called Hurwitz $\tau$-function.

The genus expansion for colored HOMFLY is also interesting, because it separates the representation- and
knot-dependencies:
\be\label{geH}
H_R^{\cal K}(q,A) &=& \left(\sigma_{[1]}^{\cal K}\right)^{|R|}\cdot
\exp \left(\sum_{\Delta} \ \hbar^{|\Delta| + l(\Delta)-2}\ S^{\cal K}_\Delta\big(A^2,\hbar^2\big)\
\varphi_R(\Delta) \right)
\ee
(in the leading order, this reproduces (\ref{spepo}), also known as exponential growth property \cite{GGS}).
The sum here goes over all Young diagrams $\Delta$, as usual,
$|\Delta|$ and $l(\Delta)$ denote respectively the number of boxes and of lines in the diagram,
$\varphi_R(\Delta)$ are the symmetric group characters: they do not depend on the knot
and are common for all Hurwitz $\tau$-functions, \cite{HurwMMS}.
Dependent on the knot are the general {\it special} polynomials
in the free energy expansion
\be
S^{\cal K}_\Delta (A^2,\hbar^2) &=&
\sum_{g\geq0} \frac{\,\hbar^{2g}}{\left(\sigma^{\cal K}_{[1]}\right)^{2g}}\cdot \sigma^{\cal K}_\Delta(g)
\ee
where $\sigma^{\cal K}_\Delta(g)$ are knot-dependent polynomials in $A$,
presumably related by the AMM/EO topological recursion for a  knot-dependent spectral curve.

For us of importance is that, since $\varphi_R(\Delta)$'s are non-diagonal, one and the same special
polynomial  $\sigma^{\cal K}_\Delta(n)$ affects the HOMFLY polynomial in different representations.
Since the lowest polynomials were already extracted from study of the symmetric representations,
we can now use them to test our new results for $R=[3,1]$.
Indeed, since $\varphi_{[3]}([2,1])=3\neq 0$ in
\be\label{genusexample}
\log \frac{H_R^{\cal K}(q,A)}{\left(\sigma^{\cal K}_{[1]}(0)\right)^{|R|}} =
\left({\hbar\over\sigma_{[1]}}\right) \cdot \sigma^{\cal K}_{[2]}(1)\varphi_R([2])
+ \left({\hbar\over\sigma_{[1]}}\right)^2\cdot\Big( \sigma^{\cal K}_{[1]}(2)\varphi_R([1]) + \sigma^{\cal K}_{[1,1]}(2)\varphi_R([1,1])
+ \sigma^{\cal K}_{[3]}(2)\varphi_R([3])  \Big) + \nn \\
+ \left({\hbar\over\sigma_{[1]}}\right)^3\cdot\Big( \sigma^{\cal K}_{[2]}(3)\varphi_R([2])
+ \sigma^{\cal K}_{[2,1]}(3)\varphi_R([2,1]) + \sigma^{\cal K}_{[4]}(3)\varphi_R([4])  \Big) + O(\hbar^4)
\ee
one can compute the first special polynomials $\sigma^{\cal K}_\Delta(g)$
with the help of only symmetric representations $R=[1],[2],[3],[4]$ using \cite{IMMMev}.

Let us briefly consider our sample example of knot $10_{161}$.
In this case,

{\footnotesize
\be
{\sigma^{10_{161}}_{[1]}(0)}  & =&   -A^6(A^4+A^2-3) \nn \\
\sigma^{10_{161}}_{[2]}(1)  & =&  2A^{12} \left( A-1 \right)  \left( A+1 \right)
\Big( 11\,{A}^{6}+8\,{A}^{4}+6\,{A}^{2}-39 \Big)
 \nn \\
\sigma^{10_{161}}_{[1]}(2)  & =&  4A^{24}\Big(A^4+A^2-3\Big)^3\Big({A}^{4}+{A}^{2}-9\Big)
\nn \\
\sigma^{10_{161}}_{[1,1]}(2)  & =&  -2A^{24}
 \Big(8\,{A}^{16}-185\,{A}^{14}+ 686\,{A}^{12}-608\,{A}^{10}+142\,{A}^{8}-1072\,{A}^{6}
+824\,{A}^{4}+1419\,{A}^{2}-1242\Big)
\nn \\
\sigma^{10_{161}}_{[3]}(2)  & =&  -2A^{24}\left( A-1 \right) ^{2} \left( A+1 \right)^{2}
\Big( 30\,{A}^{12}-319\,{A}^{10} +890\,{A}^{8}
-731\,{A}^{6}+618\,{A}^{4}-355\,{A}^{2}-348 \Big)
\nn\\
\sigma^{10_{161}}_{[2]}(3)  & =&  {4A^{36}\over3}\, \Big( A-1 \Big)  \Big( A+1 \Big)  \Big( 42\,{A}^{
22}-1975\,{A}^{20}+11805\,{A}^{18}-24834\,{A}^{16}+31718\,{A}^{14}-
32535\,{A}^{12}+\nn\\&+&10767\,{A}^{10}-2928\,{A}^{8}+23940\,{A}^{6}+37971\,{A
}^{4}-109701\,{A}^{2}+56160 \Big)
\nn
\ee
\be
\sigma^{10_{161}}_{[2,1]}(3)  & =&  {8A^{36}\over3}\, \Big( A-1 \Big)
\Big( A+1 \Big)  \Big(279\,{A}^{22}-5185\,{A}^{20}+32544\,{A}^{18}-100305\, {A}^{16}
+182258\,{A}^{14}-198924\,{A}^{12}+\nn\\&+&121074\,{A}^{10}-45282\,{A}^{8}+23637\,{A}^{6}
+8658\,{A}^{4}-40437\,{A}^{2}+22113 \Big)
 \nn \\
\sigma^{10_{161}}_{[4]}(3)  & =&  {4A^{36}\over3}\, \Big( A-1 \Big)
\Big( A+1 \Big)  \Big(660\,{A}^{22}-10753\,{A}^{20}+68289\,{A}^{18}-230418\, {A}^{16}+475094\,{A}^{14}
-627231\,{A}^{12}+\nn\\&+&545397\,{A}^{10}-331230\,{A}^{8}+147924\,{A}^{6}-33405\,{A}^{4}
-16317\,{A}^{2}+12420 \Big)
\nn  \\
\ldots
\ee
}
Now one can find the values of $\varphi_R(\Delta)$ in the table in \cite{MMN},
and, with the help of formula (\ref{genusexample}), compare the results
with the corresponding genus expansion of $H_{[3,1]}^{10_{161}}$.
They coincide, which provides yet another nontrivial check of our results for $H_{[3,1]}$.

To understand if the newly-calculated $[3,1]$-colored HOMFLY provides new special polynomials
$\sigma_\Delta$, as one could naturally expect,
a better understanding of non-linear relations between different $\varphi_R(\Delta)$ is needed.

\subsection{Differential expansion
\label{diffexpan}}

The naive genus (or loop) expansion in powers of $\hbar$ does not respect the polynomial property of HOMFLY.
This is cured in a far less trivial "differential expansion" \cite{IMMMfe,evo,arthdiff,Kondef},
which also reflects the hidden "differential structure" \cite{DGR}
lying in the base of the Khovanov approach \cite{Kho} to knot polynomials,
very different from the ${\cal R}$-matrix one exploited in the present paper.
In variance with naive genus expansion, the differential expansion
contains only a {\it finite} number of terms up to the $r+s$ power of $Z$'s,
where $Z_{I|J}^{(k)} = \{Aq^{I+k}\}\{A/q^{J-k}\}  = Z_{I+k|J-k}\sim \hbar^2$.
The expansion in the variable $h$ defined as $q=e^\hbar=(1+h)$, $A=e^{N\hbar} = (1+h)^N$ can also be made finite for each natural $N$.
What breaks the polynomiality in $h$
is the analytical continuation from integer values of $N$.
While reasonably understood for the symmetric representations,
the differential expansion remains a complete mystery beyond them, simply because
almost nothing has been known so far about the generic colored HOMFLY polynomial.
Our new results for $R=[3,1]$ allow one to make a new small step as compared to \cite{Ano21,MMMRS,MMMS21},
where only the information about $R=[2,1]$ could be used.

From \cite{Ano21} we know the $Z$-linear term of the differential expansion for the hook diagrams:
\be
H_{[r,1^{s-1}]} = 1+
\Big(Z_{2r-1|2s-1} + \sum_{i=1}^{r-1} Z_{2r-s-2i|s}+\sum_{j=1}^{s-1} Z_{r|2s-r-2j}\Big)\cdot
G_{[1]}(A,q) + O(Z^2)
\ee
with the knot-dependent polynomial $G_1(A,q)$,
which is one and the same for all representations, including the fundamental one with $r=s=1$.
Therefore, for representation $[3,1]$ with $r=3$ and $s=2$ one expects:
\be
H_{[3,1]} = 1 +
\Big(\underbrace{\{Aq^5\}\{A/q^3\} + \{Aq^3\}\{Aq\} + \{Aq^2\}\{A/q^2\} + \{A\}\{A/q^2\}}_{
Z_{5|3} + Z_{1|1}^{(2)} + Z_{3|1}^{(-1)} + Z_{1|1}^{-1}  \ \ \ \vdots \ \ \ [4]  }\cdot G_{[1]}(A,q)
\Big)
+ \{Aq^3\}\cdot O(\hbar^3)
\ee
We write down explicitly  that
the $O(Z^2)$ terms in this case are always proportional to $\{Aq^3\}$, because the
transposed $H_{[2,1,1]}$
should coincide with $H_{[1]}=1+G_1\{Aq\}\{A/q\}$ for $sl_3$ group, thus, the difference of {\it reduced}
polynomials $H_{[3,1]}-H_{[1]}=0$ at $A=q^{-3}$.
For many knots (those with the defect zero, see \cite{Kondef})
they are also proportional to $\{A\}$, but this is not the case, say, for the defect-two $8_{19}$.

There is no yet commonly accepted choice of the next terms
of the differential expansion for non-symmetric representations.
We now suggest an improved (as compared to \cite{MMMS21}) differential expansion for $[2,1]$:
\be
H_{[2,1]} = 1 + \Big(\{Aq^3\}\{A/q^3\}+\{Aq^2\}\{A\}+\{A\}\{A/q^2\}\Big)\cdot G_{[1]}(A,q)\ +
\label{diffexp21}
\ee
\vspace{-0.4cm}
$$
+ \ [3]\{Aq^2\}\{A/q^2\}\frac{\{Aq^3\}\cdot G_{[2]}(A,q) + \{A/q^3\}\cdot G_{[2]}(A,q^{-1})}{[2]}
\ +\ \{Aq^2\}\{A/q^2\}\Big(\{Aq^3\}\{A/q^3\}\cdot G_{[3]}(A,1) + \{q\}^2\cdot G_{[2,1]}(A,q)\Big)
$$
In this case, $G_{[21]}$ is of the order $\hbar^2$ like $G_{[3]}$
(we remind that the symmetric representation coefficients are always $G_{[r]}\sim \hbar^{r-1}$,
but they are explicitly divisible by factors $\{Aq^i\}$ only for small {\it defects}, see \cite{Kondef}).
We remind  that $G_{[21]}$ drops out of the differential expansion for the special polynomial (i.e. at $q=1$).
Also
\be
G_{[2]}(A,q^{-1})= G_{[2]}(A,q^{-1})=G_{[1,1]}(A,q)
\ \ \ \ \ \ \  {\rm and} \ \ \ \ \ \ \
G_{[2,1]}(A,q^{-1}) = G_{[2,1]}(A,q)
\ee
what makes the expression explicitly symmetric
under the transposition of Young diagram.

Searching for a counterpart of (\ref{diffexp21}) for $R=[31]$, we use
the three properties:
\be
H_{[3,1]}-H_{[2]}\sim \{A/q^2\} & \Longrightarrow & H_{[3,1]}\ \stackrel{A=q^2}{=}\ H_{[2]}  \nn \\
H_{[3,1]}-H_{[1]}\sim \{Aq^3\} & \Longrightarrow & H_{[2,1,1]}\ \stackrel{A=q^3}{=}\ H_{[1]} \nn \\
H_{[3,1]}-H_{[2]}\sim \{Aq^3\} & \Longrightarrow & H_{[2,1,1]}\ \stackrel{A=q^3}{=}\ H_{[1,1]}
\ee
which implies that
\be
H_{[3,1]} = 1 +
\Big( \{Aq^5\}\{A/q^3\} + \{Aq^3\}\{Aq\} + \{Aq^2\}\{A/q^2\} + \{A\}\{A/q^2\} \Big)\cdot G_{[1]}(A,q) + \nn \\
+\{Aq^3\}\{Aq^2\}\{A/q\}\cdot G_2(A,q) + \ \{Aq^3\}\{A/q^2\}\cdot \Big({\rm something}\Big)
\label{fact31}
\ee
"Something" should complement this expression so that it contains six terms of the order of $G_2$,
four terms of the order of $G_3$ and one term of the order of $G_4$, with possible corrections by $G_{21}$
and $G_{31}$.
Eq.(\ref{fact31}) is true for all our examples, but this only checks their consistency with
representation theory.
Less trivial (but more speculative) conjectures about the non-symmetric differential expansion
will be presented elsewhere.

Since we devoted a piece of this paper to implications of the evolution method, it deserves mentioning that
description of the knot polynomials in terms of the differential expansion parameters $G_R$
is badly consistent with the evolution considered in s.\ref{evoser}.
This is clear already from the fact that unity, the first term in the
differential expansion is {\it not} an eigenvalue of the ${\cal R}$-matrix
in the $R^{\otimes 2}$ channel (it {\it is} in the channel $R\otimes \bar R$, and this
explains a nice consistency of the evolution and differential expansion for
the peculiar family of {\it twist} knots \cite{IMMMfe,evo,twist}).
In general the interplay between the two hidden structures,
the evolution and differential expansion coefficients leads to a very interesting and
important puzzle in knot theory.

\section{Conclusion \label{conc}}

In this paper we report the results
of tedious calculation of the Racah (mixing)
${\cal U}_Q$ matrices for $[3,1]^{\otimes 3} \longrightarrow Q$
and their application to study of the 3-strand knots and links.
These results confirm previous conjectures and slightly extend
our understanding of complicated issues
like evolution method \cite{DMMSS,evo} and
differential expansion \cite{DGR,arthdiff}.
This advance was made possible by application of the
highest weight method, which we developed in \cite{MMMS21}.
The size of this paper does not reflect the actual complexity and amount
of calculations, we just commented briefly on
various problems encountered and solved on the way,
and provided simple illustrations for the results at different steps
of calculations.
For the full set of newly-derived  {\bf mixing matrices}
(additionally converted to the block form)
see \cite{knotebook}: they are too big for a paper,
but are nicely handled by the eigenvalue hypothesis \cite{IMMMev}.
Also at \cite{knotebook} there are the {\bf $[3,1]$-colored HOMFLY polynomials} for
the 3-strand knots up to 10 crossings  and  the {\bf evolution coefficients} for the entire
infinite next-to-twist {\it family} $(n,-1\,|\,1,1) = \{(n+3)_2,(2-n)_3\}$.
In this paper, we also updated {\bf the list of properties} in sec.\ref{proppols},
including {\bf the improved differential expansion} for asymmetric representations.

This list of 5 boldfaced items describes the main results reported in the present paper.

\bigskip

The next steps in study could be:

\bigskip

$\bullet$
Finding the inclusive Racah matrices $R^{\otimes 3} \longrightarrow all$ for higher representations $R$.
This can look like impossible dream, but there is certain evidence that general
formulas can exist.
Still, this requires an essential progress in the highest weight method,
we devote a separate publication to these perspectives.

\bigskip

$\bullet$ Finding the Racah matrices ${\cal S}_R$
for $R\otimes \bar R\otimes \bar R \longrightarrow R$
and $\bar{\cal S}_R$ for $R\otimes R\otimes \bar R \longrightarrow R$.
These are much simpler, because only one outgoing representation $Q=R$ is requested.
They are more complicated for the highest weight method, because $\bar R$
depends on $N$ (of $sl_N$) and so do the highest weights.
As explained in \cite{inds,MMMRS,MMMRSS},
the knowledge of these matrices allows one to handle all arborescent (double fat) knots \cite{Con,arbor},
and merging that knowledge with the results of the present paper extends this
to {\it the fingered 3-strand knots} \cite{MMfing,MMMS21,MMMRSS} which include
at least the entire Rolfsen table up to $10$ intersections and, after applying the power
{\it family} method from these papers, many more.
The Racah matrices ${\cal S}$ and $\bar{\cal S}$ for the symmetric representations
are well known \cite{Racah,gmmms,MMSpret}, for $R={[2,1]}$ they were found in
\cite{GJ}.

\bigskip

$\bullet$
In fact, a simple trick allows one to extract ${\cal S}$
from comparison of two different expressions, 3-strand and Pretzel ones
for a peculiar two-parametric evolution family $(m,-1|n,-1)={\rm Pr}(m,n,\bar 2)$.
The second matrix $\bar{\cal S}$ can then be built from ${\cal S}$ by equation (63) from \cite{MMMRS}.
The result, requiring a generalization of our analysis in sec.5, will be
presented elsewhere \cite{mmmsSS}.

\bigskip

$\bullet$ A serious shortcut to the Racah matrix calculus can be provided
by the eigenvalue hypothesis of \cite{IMMMev}, see also \cite{univev},
expressing these matrices through the much simpler eigenvalues of
quantum ${\cal R}$-matrices (for which there is a general formula through the
eigenvalues of the Casimir/cut-and-join-operator \cite{MMN,DMMSS}).
Even if the very hypothesis is true,
it suffers from two uncertainties: the sign choice
(well defined are the squares of matrix elements)
and the problem of degenerate ${\cal R}$-matrix eigenvalues.
One of the results of our studies in this paper is a confirmation
of a natural assumption: that this degeneracy just implies
a decomposition of mixing matrices into simpler (smaller-size) blocks,
which are provided by the same eigenvalue hypothesis.
Further work in this direction looks quite promising,
and, perhaps, can provide general formulas for the Racah matrices
much sooner than any alternative approach.

\section*{Acknowledgements}

This work was funded by the Russian Science Foundation (Grant No.16-12-10344).

\end{document}